\documentclass{emulateapj}
\usepackage{amssymb}
\usepackage{natbib}
\usepackage{epsfig}
\usepackage{aas_macros}
\usepackage{lscape}
\usepackage{longtable}

\newcommand{\Yx}{\ensuremath{Y_{\mathrm{X}}}}

\newcommand{\Msol}{\ensuremath{\mathrm{M_{\odot}}}}

\newcommand{\rt}{\ensuremath{R_{\mathrm{200}}}}
\newcommand{\rf}{\ensuremath{R_{\mathrm{500}}}}
\newcommand{\rn}[1]{\ensuremath{R_{\mathrm{#1}}}}

\newcommand{\rd}{\ensuremath{r_{\mathrm{d}}}}

\newcommand{\Zsol}{\ensuremath{\mathrm{Z_{\odot}}}}
\newcommand{\fgas}{\ensuremath{f_{\mathrm{gas}}}}

\newcommand{\OM}{\ensuremath{\Omega_{\mathrm{M}}}}

\newcommand{\zo}{\ensuremath{z_{\mathrm{obs}}}}
\newcommand{\zf}{\ensuremath{z_{\mathrm{form}}}}

\newcommand{\egc}{{\it e.g.}}  
\newcommand{\etal}{{\it et al.\thinspace}}

\newcommand{\Chandra}{\emph{Chandra}}

\newcommand{\ROSAT}{\emph{ROSAT}}
\newcommand{\XMM}{\emph{XMM-Newton}}

\newcommand{\MEKAL}{\textsc{MeKaL}}

\newcommand{\chisq}{\ensuremath{\chi^2}}

\newcommand{\gta}{\,\rlap{\raise 0.4ex\hbox{$>$}}{\lower 0.6ex\hbox{$\sim$}}\,}  
\newcommand{\lta}{\,\rlap{\raise 0.4ex\hbox{$<$}}{\lower 0.6ex\hbox{$\sim$}}\,}  

\newcommand{\nm}{\mbox{\ensuremath{\mathrm{~\nm}}}}

\newcommand{\cm}{\mbox{\ensuremath{\mathrm{~cm}}}}

\newcommand{\km}{\mbox{\ensuremath{\mathrm{~km}}}}

\newcommand{\kpc}{\mbox{\ensuremath{\mathrm{~kpc}}}}
\newcommand{\Mpc}{\mbox{\ensuremath{\mathrm{~Mpc}}}}
\newcommand{\s}{\mbox{\ensuremath{\mathrm{~s}}}}
\newcommand{\ks}{\mbox{\ensuremath{\mathrm{~ks}}}}

\newcommand{\yr}{\mbox{\ensuremath{\mathrm{~yr}}}}
\newcommand{\Gyr}{\mbox{\ensuremath{\mathrm{~Gyr}}}}

\newcommand{\keV}{\mbox{\ensuremath{\mathrm{~keV}}}}
\newcommand{\erg}{\mbox{\ensuremath{\mathrm{~erg}}}}

\newcommand{\arcm}{\ensuremath{\mathrm{^\prime}}}
\newcommand{\arcs}{\arcm\hskip -0.1em\arcm}

\newcommand{\K}{\mbox{\ensuremath{\mathrm{~K}}}}

\newcommand{\cmsq}{\ensuremath{\mathrm{\cm^2}}}

\newcommand{\pcc}{\ensuremath{\mathrm{\cm^{-3}}}}

\newcommand{\pcmsq}{\mbox{\ensuremath{\mathrm{~cm^{-2}}}}}
\newcommand{\pMpc}{\ensuremath{\mathrm{\Mpc^{-1}}}}
\newcommand{\ps}{\ensuremath{\mathrm{\s^{-1}}}}

\newcommand{\ergps}{\ensuremath{\mathrm{\erg \ps}}}

\newcommand{\ent}{\ensuremath{\mathrm{\keV \cmsq}}}

\newcommand{\kmpspMpc}{\ensuremath{\mathrm{\km \ps \pMpc\,}}}
\newcommand{\LT}{\mbox{\ensuremath{\mathrm{L-T}}}}

\newcommand{\MT}{\mbox{\ensuremath{\mathrm{M-T}}}}

\newcommand{\YM}{\mbox{\ensuremath{\mathrm{Y_X-M_{500}}}}}

\newcommand{\LCDM}{$\Lambda$CDM~}

\newcommand{\jjj}{ClJ1226.9$+$3332}

\begin{document}


\title{Deep \XMM\ and \Chandra\ observations of ClJ1226.9$+$3332: A
detailed X-ray mass analysis of a $z=0.89$ galaxy cluster.}

\author{B. J. Maughan\altaffilmark{1}$^,$\altaffilmark{$\dagger$}}
\author{C. Jones\altaffilmark{1}}
\author{L. R. Jones\altaffilmark{2}}
\author{L. Van Speybroeck\altaffilmark{1}}
\altaffiltext{1}{Harvard-Smithsonian Center for Astrophysics, 60 Garden St, Cambridge, MA 02140, USA.}
\altaffiltext{2}{School of Physics and Astronomy, The University of Birmingham,
  Edgbaston, Birmingham B15 2TT, UK}
\altaffiltext{$\dagger$}{\Chandra\ fellow}
\email{bmaughan@cfa.harvard.edu}

\shorttitle{A detailed X-ray mass analysis of \jjj.}

\shortauthors{B. J. Maughan \etal}


\begin{abstract}
Deep \XMM\ and \Chandra\ observations of \jjj\ at $z=0.89$ have enabled the most
detailed X-ray mass analysis of any such high-redshift galaxy cluster. The
\XMM\ temperature profile of the system shows no sign of central cooling, with a
hot core and a radially declining profile. A temperature map shows
asymmetry with a hot region that appears to be associated with a subclump
of galaxies at the cluster redshift, but is not visible in the X-ray
surface brightness. This is likely to be result of a merger event in the
cluster, but does not appear to significantly affect the overall
temperature profile. The \XMM\ temperature profile, and combined \Chandra\
and \XMM\ emissivity profile allowed precise measurements of the global
properties of \jjj; we find $kT=10.4\pm0.6\keV$, $Z=0.16\pm0.05\Zsol$, and
$M=5.2^{+1.0}_{-0.8}\times10^{14}\Msol$. We obtain profiles of the
metallicity, entropy, cooling time and gas fraction, and find a high
concentration parameter for the total density profile of the system. The
global properties are compared with the local \LT\ and \MT\ relations, and
we are able to make the first observational test of the predicted evolution
of the \YM\ relation. We find that departures from these scaling relations
are most likely caused by an underestimate of the total mass by $\sim30\%$
in the X-ray hydrostatic mass analysis due to the apparent recent or
ongoing merger activity.
\end{abstract}


\section{Introduction}
Clusters of galaxies are an important tool for testing and improving
cosmological models. High-redshift clusters, in particular, provide the
strongest constraints. Measurements of the total and baryonic mass
functions, and baryon fractions of distant clusters have been used to
measure cosmological parameters \citep[\egc][]{vik03,hen04,all04}. Even
with the precise measurements of cosmological parameters provided by cosmic
microwave background and supernova observations
\citep[\egc][]{spe03,ton03}, clusters play an important role by providing
independent constraints with different parameter degeneracies
\citep{all04}. Large samples of distant clusters detected in planned X-ray
and Sunyaev-Zel'dovich effect (SZE) surveys will be able to place strong
constraints on the dark energy density and equation of state, but the most
powerful constraints require mass estimates for many clusters
\citep{maj03}.

The most reliable X-ray mass estimates require radial temperature and
density profiles of the intra-cluster gas to allow the solution of the
equation of hydrostatic equilibrium. However, the low X-ray fluxes of
typical high-redshift clusters make the observing time needed to meet these
requirements prohibitive. Sufficiently deep observations for clusters at
$z>0.5$ are very rare \citep[\egc][]{jel01,arn02b,don03,mau04b}; for most
clusters at high redshifts, masses must be estimated from observed global
properties such as X-ray luminosity and temperature. The scaling relations
between cluster properties at high redshifts are then of key importance,
and their uncertainty is a dominant contributor to the error budget of
derived cosmological parameters \citep{hen04}. Furthermore, studies of the
evolution of the cluster scaling relations can be used to probe cluster
formation and the history of non-gravitational heating
\citep[\egc][]{ett03,mau06a}.

At $z=0.89$, galaxy cluster \jjj\ is the hottest, most
luminous known at $z>0.6$.\footnote{Comparisons here are made with all
clusters listed in the BAX database (\url{http://bax.ast.obs-mip.fr}), a compilation of published X-ray properties of galaxy
clusters} The only comparable system is MS1054.5$-$0321 at $z=0.83$, an
actively merging system which is somewhat cooler and less luminous than
\jjj\ \citep{jee05b}. Unlike that very disturbed system, \jjj\ has a relaxed
morphology, making it a promising candidate for a reliable hydrostatic mass
estimate. \jjj\ was detected in the Wide Angle \ROSAT\ Pointed Survey
\citep[WARPS][]{sch97,ebe01a} and an analysis of a $15\ks$ AO1 \XMM\
observation confirmed its temperature at $11.5\keV$ with a bolometric
luminosity of $5\times10^{45}\ergps$ and a relaxed morphology
\citep{mau04a}. An early, short \Chandra\ observation of \jjj\ was
presented by \citet{cag01}, who also pointed out that the VLA FIRST
catalogue shows a faint point-like radio source at the cluster centre,
consistent with a low-luminosity radio-loud AGN. \jjj\ has also been
detected in the SZE \citep{joy01} and the properties of the member galaxies
have been investigated \citep{ell04,ell06}.

In this {\it paper} we present new \XMM\ and \Chandra\ observations of
\jjj. In the following sections we describe the data reduction and
background subtraction techniques, and present the results of the imaging
and spectral analysis. The radial profiles of the clusters properties such
as mass, temperature and entropy are then investigated and the global
properties of the system are compared with those of other clusters. In the
appendix we present a novel method for defining contour levels in smoothed
X-ray images. Numerical subscripts are used on cluster properties to
indicate the overdensity with respect to the critical density at
the radius at which (or within which) the property was measured. The
critical density at the cluster's redshift is used in this
definition. We assume a flat \LCDM\ cosmology with $H_0=70\kmpspMpc$ and
$\OM=0.3$ throughout, and uncertainties are quoted at the $68\%$ level.

\section{Data Reduction}

The \XMM\ data (ObsID 0200340101) were reduced and analysed with the
Science Analysis Software (SAS) version 6.5, and the most recent
calibration products available as of 15 November 3005. A lightcurve of the
field of view was produced in the $10-15\keV$ band for each of the three
EPIC cameras. The lightcurves were filtered to reject periods of high
background, and approximately $16\ks$ of data were rejected. The cleaned
lightcurves exhibited stable mean count rates and the clean exposure times
were $68\ks$ and $75\ks$ for the PN and MOS detectors respectively. A local
background region, and the \citet{rea03} blank-sky background datasets were used for
background estimates where appropriate, as discussed below.

The \Chandra\ data consisted of two ACIS-I observations (ObsID 3180 and
5014) which were were reduced and analysed consistently. The level 1 events
files were reprocessed with the latest calibration (as of November 2005),
applying the charge-transfer inefficiency and time-dependent gain
corrections. Background flare filtering was performed with {\it lc\_clean}
in the $0.3-12\keV$ band excluding the target CCD, and all sources detected
by the pipeline processing. The good time remaining for each observation
was $25\ks$. The period D blank-sky background files \citep{mar00b} were
appropriate for both observations, and these were reprojected onto the sky
coordinates of each observation. Both observations were taken in VFAINT
telemetry mode, so the additional VFAINT cleaning procedure was applied to
the source and background
datasets\footnote{http://asc.harvard.edu/ciao/threads/aciscleanvf}.

\section{Background Considerations}\label{s.bgspec}
In periods free of background flaring, the background is dominated by
X-rays at low energies ($\lta2\keV$) and particles at high energies
($\gta2\keV$). The X-ray component includes a significant contribution from
Galactic emission which varies with position on the sky. The blank-sky background
datasets are compiled from multiple pointings with different Galactic X-ray
fluxes, so may not agree well with the soft X-ray background in the source
dataset. However, the blank-sky datasets enable background estimates to be
made at the same detector position as the source, eliminating instrumental
effects, and furthermore, many extended sources fill the field of view of
\Chandra\ or \XMM\, prohibiting the measurement of the background in the
source dataset. Due to its high redshift, \jjj\ occupies a small area on
the detectors, leaving large source-free regions in which the background
emission can be measured. This local background was compared with the
blank-sky files to determine the optimal choice of background for imaging
and spectral analyses.

Images were created in the $0.7-2\keV$ band for each
detector, along with exposure maps to correct for chip gaps, bad pixels,
and telescope vignetting. The three \XMM\ PN and MOS images and the two
\Chandra\ ACIS-I images were combined to give separate \XMM\ and \Chandra\
images, and these were exposure corrected. Background images and exposure
maps were created in the same way from the blank-sky datasets for each
detector. Sources in these images were then detected with the wavelet
detection algorithm of \citet{vik98b}. The detected point and extended
sources were then excluded from the analysis of \jjj.

Due to the possible difference in the soft Galactic X-ray flux between the source
and background datasets, the blank-sky images were
normalised to match the flux in the source-free regions in the target
images. For both the \XMM\ and \Chandra\ images, radial profiles of the
source and background images and their exposure maps were created and used
to determine the detection radius (\rd) of the cluster emission. Emission
within \rd\ was then excluded and the background images were normalised to
the remaining emission. The detection radius was the recomputed and the
process was repeated until \rd\ converged. This procedure provides a
reliable measurement of both the cluster extent and the normalisation of
the blank-sky images in the imaging band. 

\begin{figure}
\begin{center}
\scalebox{0.35}{\includegraphics*[angle=270]{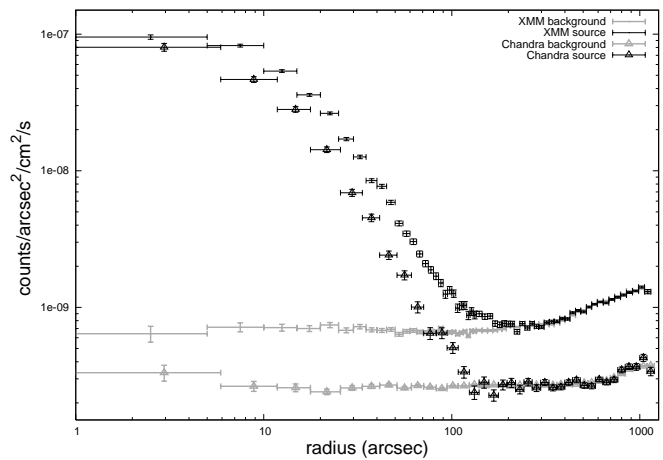}} \\
\caption[]{\label{f.rprofs}Source and background radial surface brightness profiles of \jjj\ 
measured with \XMM\ and \Chandra.}
\end{center}
\end{figure}

Fig. \ref{f.rprofs} shows the source and background profiles for the
\XMM\ and \Chandra\ images. The good agreement of the source and background
profiles at $r>300\arcs$ indicates that the source exclusion and background
normalisation were successful. Note that the increasing background with
radius (most noticeable in the \XMM\ data) is due to the fact that the
particle background component is not vignetted, so the exposure correction
incorrectly boosts this background component. In this imaging analysis, the
effect is the same in the source and background images and cancels
out. However, this effect is a cause for concern if an off-axis region is
used to extract a local background spectrum. These normalised blank-sky
background images were thus used for all of the imaging analysis for all
datasets, while the background subtraction for the spectral analysis is
discussed in more detail below. The consistency of the \XMM\ and
\Chandra\ surface brightness profiles of \jjj\ is demonstrated in section
\textsection \ref{s.gdens}.

\begin{figure*}
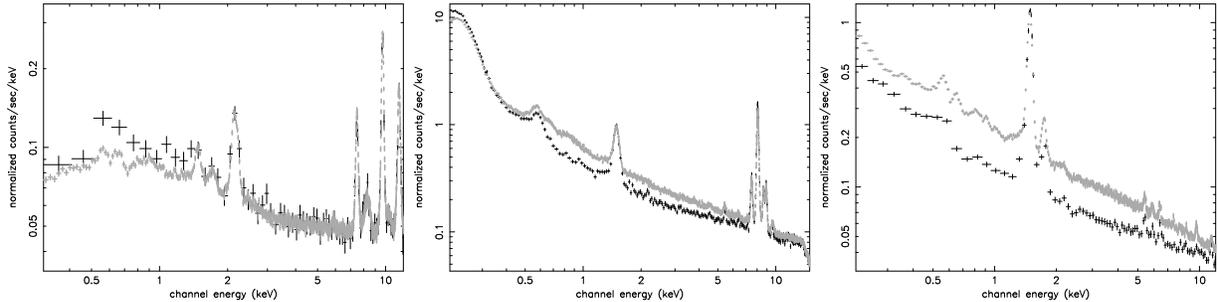

\begin{center}
\includegraphics[angle=-90,width=5.3cm]{3180bgspec_bw.ps}
\includegraphics[angle=-90,width=5.3cm]{j1226bgallpnbgspec_bw.ps}
\includegraphics[angle=-90,width=5.3cm]{j1226bgallm1bgspec_bw.ps}\\
\caption{\label{f.bgspec}Background spectra taken from the target (black
points) and blank-sky (grey points) datasets for the \Chandra\ ACIS-I ObsID
3180 (left), \XMM\ PN (centre) and \XMM\ MOS (right) observations of \jjj.}
\end{center}
\end{figure*}

A prime motivation in defining \rd\ was to fully exclude any significant
cluster emission to enable to measurement of the background in the target
observations. To this end \rd\ was conservatively defined as the
radius beyond which no further radial bins had a detection
of $>0.5\sigma$ above the background. The detection radius for \XMM\ was
$210\arcs$ and for \Chandra\ was $126\arcs$. For each detector, spectra
were extracted from the target datasets after excluding all sources, and
the cluster emission out to \rd, to give a local background spectrum. Blank
sky background spectra were extracted for each
detector with the same spatial regions excluded, and the local and blank
sky background spectra are plotted in Fig. \ref{f.bgspec}.

The \Chandra\ blank-sky background spectra were normalised to match the
$9.5-12\keV$ count rates in the target datasets. At these high energies,
the background is dominated by particle events. The left panel of
Fig. \ref{f.bgspec} shows the background spectra for the \Chandra\ ObsID
3180 observation (ObsID 5014 is consistent). The good agreement between the
spectra above $1.5\keV$ indicates that relative contributions of the
particle and X-ray background components is the same in the target and
blank-sky datasets. The excess in the local background at
the soft end is due to Galactic X-ray emission. These soft residuals are
accounted for in our spectral analysis of the \Chandra\ data by including a
model for this soft thermal component (see \textsection \ref{s.spec}). We
conclude that the blank-sky background files provide a reliable
background measurement for the \Chandra\ datasets.

The \XMM\ blank-sky background spectra were normalised by the ratio of the
count rates of events detected outside the field of view of the \XMM\
telescopes. These events are due solely to the particle background. It is
clear in the centre and right panels of Fig. \ref{f.bgspec} that the flux
in the particle-induced fluorescent lines agrees well in the target and
blank-sky spectra, but the flux elsewhere is significantly lower in the
local background spectra. This suggests that the
relative contribution of the particle background component is significantly
greater in the target dataset. A similar effect has been found in other
\XMM\ observations \citep{mau06b,kho06}, and we emphasise the importance of
comparing the blank-sky and local background spectra whenever possible. A
possible resolution to this discrepancy would be to
normalise the spectra by the count rate in the $2-7\keV$ band, but this
would lead to an underestimate of the flux in the fluorescent
lines. Furthermore, the slopes of the of the spectra are different due to
the differing particle background contributions. We thus conclude that the
blank-sky datasets cannot be used for the spectral analysis of this \XMM\
observation. Instead, we define a local background annulus region between
$250<r<500\arcs$. 

This local background region lies further off-axis than the cluster
emission, so the vignetting of the background photons must be
considered. Vignetting effects were corrected by using the SAS task {\it
evigweight}. This assigns a weight to every event based on the ratio of the
effective area at its detected position and energy, to the effective area
on-axis at the same energy. This procedure can not distinguish between
vignetted X-rays, and non-vignetted particle events, leading to the
incorrect boosting of the particle background. Fig. \ref{f.rprofs}
demonstrates that this is already a significant effect by $r\sim400\arcs$.

For a given source region, this effect can be removed with the aid of the
\XMM\ closed filter background datasets which consist of purely instrumental and particle
events. A ``boosted-particle'' spectrum was created for each camera by
running evigweight on the closed filter background and extracting a
spectrum from the local background region. This replicates the treatment of
the particle events in the source dataset. These boosted-particle spectra
were then normalised to the $10-14\keV$ count rate in the local background
spectra and subtracted from them, creating a vignetting-corrected ``X-ray only''
background spectrum. A similar boosted-particle spectrum was then extracted
from the {\it source} region in the closed filter data, normalised by the
same factor and added onto the ``X-ray only'' background spectrum (after
scaling by area). The resultant background spectrum then consisted of an
X-ray background that was correctly weighted for the vignetting at the
background region, and a particle component that was incorrectly weighted
in exactly the same way as the particles in the source spectrum. The net
effect is that the particle and X-ray backgrounds are both subtracted
correctly. This has a significant effect on the temperatures measured
at large radii where the background dominates, and this method was followed
for all \XMM\ spectral analysis.

\section{Cluster Morphology}\label{s.morph}
The exposure corrected \Chandra\ and \XMM\ images of \jjj\ were smoothed to
the $3\sigma$ level using the adaptive smoothing algorithm of
\citet{ebe06a}. Contours were then derived from the smoothed images at flux
levels defined so that the emission in the region between two adjacent contour
levels was detected at $>3\sigma$ above the emission in the next lowest
band. This contouring algorithm is described in detail in Appendix
\ref{s.con} and allows the significance of features in contour plots to be
determined easily by eye.

\begin{figure}
\begin{center}
\plotone{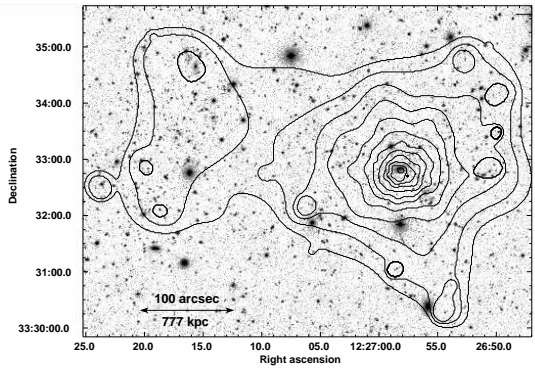}
\caption{\label{f.xmmcon}Contours of the emission detected by \XMM\ are
overlaid on a Keck-II I band image. The foreground cluster ClJ1227.3$+$3333
($z=0.766$) is visible in the eastern part of the image.} 
\end{center}
\end{figure}

\begin{figure}
\begin{center}
\plotone{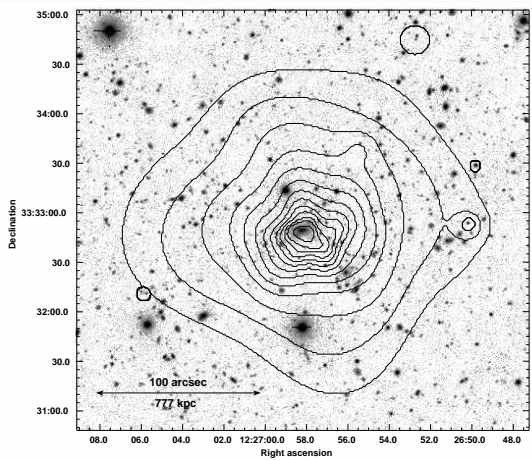}
\caption{\label{f.chandracon}Contours of the emission detected by \Chandra\ are
overlaid on a Keck-II I band image.} 
\end{center}
\end{figure}

Fig. \ref{f.xmmcon} shows a wide field view of \jjj\ with \XMM\ contours
overlaid on a Keck-II I band image. The image is dominated by the bright,
circular \jjj\ in the western half, but the foreground cluster
ClJ1227.3$+$3333 is visible in the eastern half of the
image. ClJ1227.3$+$3333 appears unrelaxed and consists of several
sub-clumps of galaxies at a similar spectroscopic redshift ($z=0.766$). All
of the sources seen in the contours are consistent with point sources except for the
northern peak of ClJ1227.3$+$3333 and
\jjj. Fig. \ref{f.chandracon} shows the \Chandra\ contours of \jjj\ on the
same optical image. These contours demonstrate the regular, relaxed
appearance of the emission in \jjj.

\section{Spectral Analysis}\label{s.spec}
\subsection{Methodology}
Before discussing the different spectral analyses performed, we give an
overview of the spectral extraction and fitting methods used for the \XMM\
and \Chandra\ data. In all cases, responses were generated for the spectral
extraction regions, weighted by the distribution of counts in the
region. Cluster spectra were modelled in XSPEC with a single temperature,
APEC \citep{smi01} plasma model, and with the exception of the MOS data as
noted below, the absorbing column was fixed at the Galactic value
inferred from $21\cm$ observations \citep{dic90}. During fitting, the APEC
normalisation, metal abundance and temperature were the only free
parameters. When data from multiple detectors were fit simultaneously, the
temperature and metal abundances of the models were tied, but the
normalisations were not.

For the \XMM\ data, only the best-calibrated events (FLAG=0) were
selected, and the modified local background spectra discussed in
\textsection \ref{s.bgspec} were used for each detector. The MOS and PN spectra
were fit in the $0.4-7\keV$ energy bands,
avoiding the bright fluorescent lines above $7\keV$ in the PN and
minimising any remaining problems with the particle background subtraction.

For the \Chandra\ data, the background spectra were taken from the blank
sky datasets. A spectrum of the soft residuals due to Galactic X-ray
emission differences between the target and blank-sky datasets was produced
for each dataset by subtracting a blank-sky background spectrum from a
local background spectrum extracted in the same region. This was fit with a
$0.18\keV$, unabsorbed APEC model of zero redshift \citep{vik05a}. This model was
included as an additional background component in all spectral fits, with
the normalisation scaled by the difference in extraction area. The cluster
spectrum was fit by an absorbed APEC model in the $0.6-9\keV$ and the best
fitting temperature was found. The model was refit with the soft Galactic
component normalisation set to $\pm1\sigma$ and again with the overall
blank-sky background renormalised by $\pm2\%$, and the resulting systematic
temperature uncertainties were added in quadrature to the statistical
uncertainties.

\subsection{Cross-calibration}\label{s.cal}
\begin{figure}
\begin{center}
\includegraphics*[angle=270,width=7.5cm]{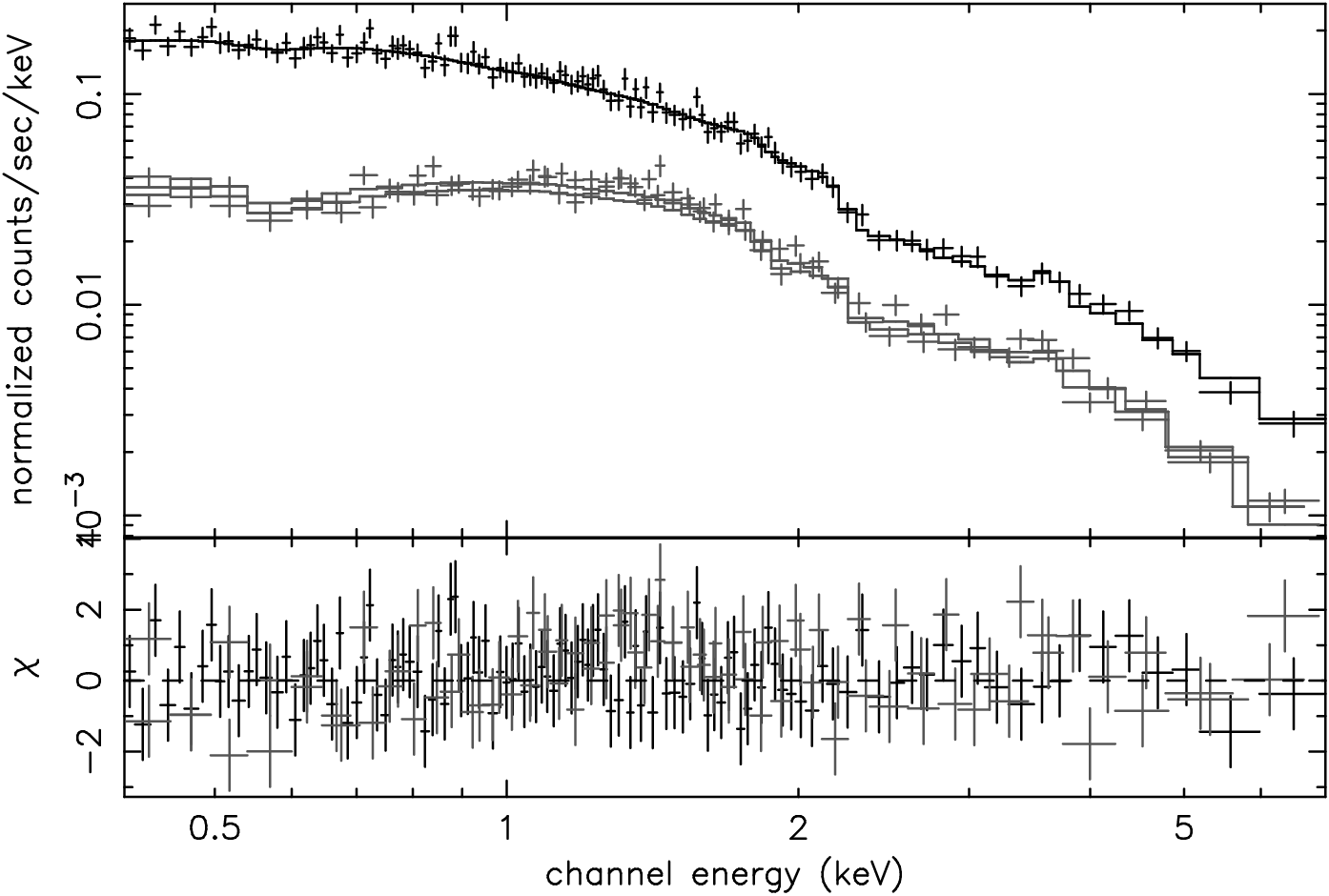}
\hspace{0.5cm}
\includegraphics*[angle=270,width=7.5cm]{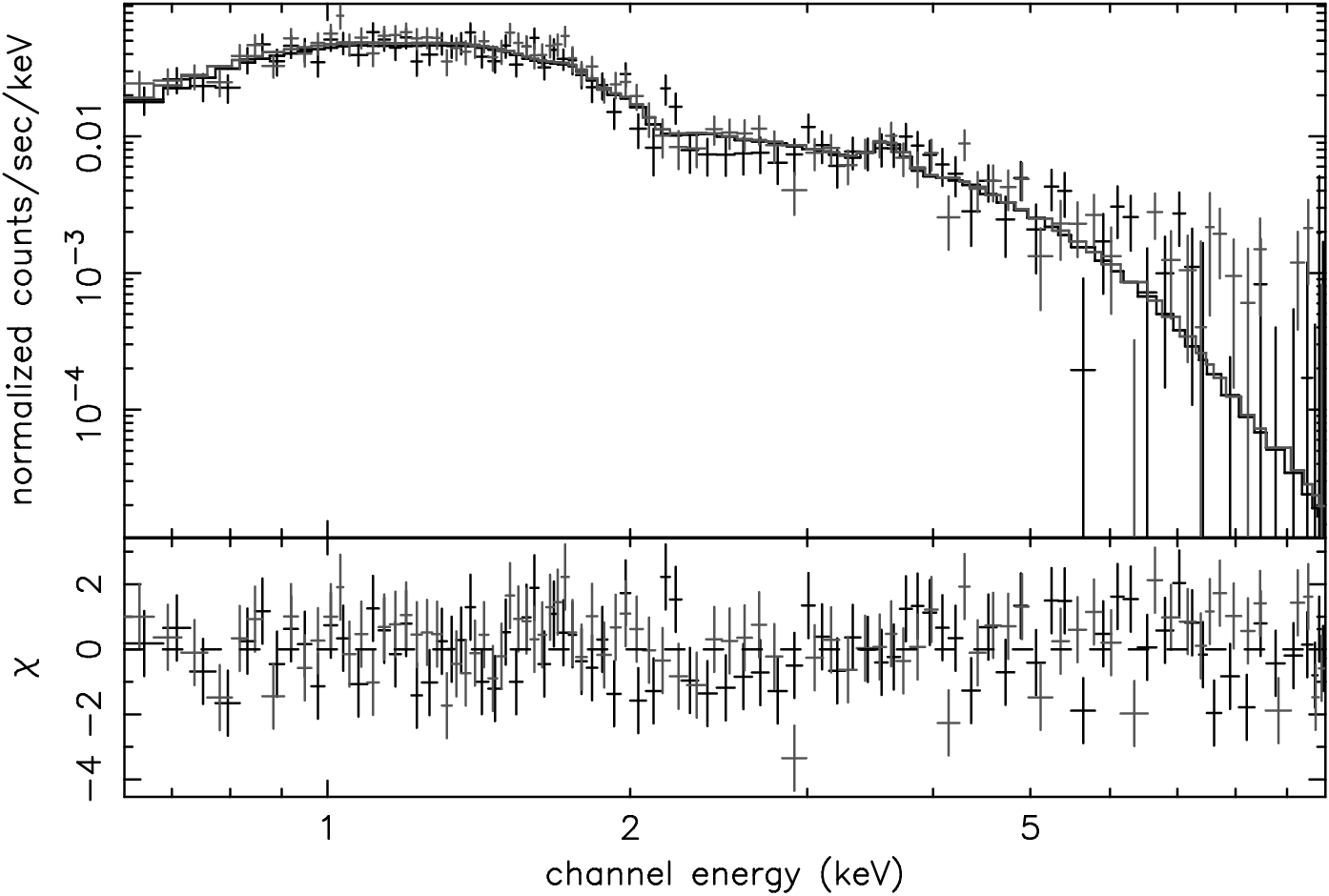} \\
\caption[]{\label{f.spec}Spectra extracted from the central $50\arcs$ of
\jjj\ plotted with best fitting models with $Z=0.3\Zsol$. {\it Top:} \XMM\
spectra; the temperature of the 3 models were tied together, and the
obsorbing column was fixed at the Galactic value for the PN data and at
$4.1\times10^{20}\pcmsq$ for the MOS data (lower lines). {\it Bottom:} \Chandra\ spectra
with soft residuals modelled simultaneously. The temperatures of the
\Chandra\ cluster spectra were tied together.}
\end{center}
\end{figure}

In order to test the spectral calibration of the different detectors before
combining their data, a single spectrum was extracted from $r<50\arcs$ for
each detector. Selecting only the central cluster region in this way
maximised the signal to noise ratio of the spectra, reducing the
uncertainties due to the background subtraction. No
point sources were detected within this region by either \Chandra\ or
\XMM. To further simplify the comparison, the metal abundances were fixed at
$0.3\Zsol$. The spectra were fit separately and simultaneously in different
combinations, and example spectra are plotted in Fig. \ref{f.spec}. 

\begin{figure}
\begin{center}
\scalebox{0.35}{\includegraphics*{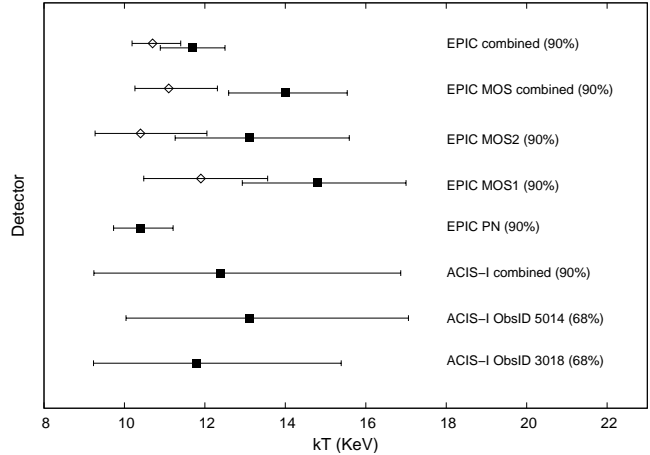}}
\caption[]{\label{f.cal}Comparisons of temperatures measured for the
central $50\arcs$ of \jjj\
with \XMM\ and \Chandra\ detectors. The labels indicate the detector
combination and confidence level for each measurement. The \XMM\ spectra
were fit in the $0.4-7\keV$ band. Solid points indicate temperatures
measured with the absorbing column fixed at the Galactic value; for the
hollow points the absorbing column for the MOS detectors was fixed at $4.1\times10^{20}\pcmsq$.}
\end{center}
\end{figure}

The results of the spectral fitting are summarised in Fig. \ref{f.cal}. The
solid points show that the temperatures measured with EPIC MOS were
significantly higher than the PN measurements; the $90\%$ confidence
intervals do not overlap. The absorbing column was then allowed to vary,
and the agreement in $kT$ was improved dramatically. The best fitting PN
absorbing column did not vary significantly from the Galactic value
($1.38\times10^{20}\pcmsq$) and the PN temperature was unchanged. In the
case of the combined MOS spectra, however, the best-fitting absorption was
significantly higher than the Galactic value at
$4.1\pm1.0\times10^{20}\pcmsq$ ($90\%$ errors) and the temperature was
significantly reduced. The absorbing columns found when fitting the MOS
data independently of each other were also consistent with this
value. All of these spectral fits were statistically acceptable, whether
the absorbing column was fixed or varying.

The hollow points in Fig. \ref{f.cal} show the temperatures measured when
the absorbing column was fixed at $4.1\times10^{20}\pcmsq$ for the MOS
data; all EPIC measurements agree well. The statistical precision of the
ACIS measurements, meanwhile, cannot exclude any of the
\XMM\ measurements. The fact that the MOS fits require an absorbing column
significantly higher than Galactic suggests that there is an outstanding
calibration problem with those detectors, and that the PN temperature
should be more reliable. In all further spectral fits, the absorbing column
was frozen at the Galactic value for the PN and at $4.1\times10^{20}\pcmsq$
for the MOS data. The variation in the best-fitting temperature caused by setting
the MOS absorbing column to its $90\%$ limits was taken as an indication of
the additional systematic uncertainty due to this problem. This uncertainty
was added in quadrature to the statistical errors on all subsequent
temperature measurements, resulting in an increase of
$\approx50\%$. Finally, we note that consistent temperatures and
uncertainties were found using the combined EPIC data if energies below
$1.2\keV$ were ignored in the MOS spectra.

\subsection{Spectral Profiles}
\begin{figure}
\begin{center}
\scalebox{0.35}{\includegraphics*[angle=270]{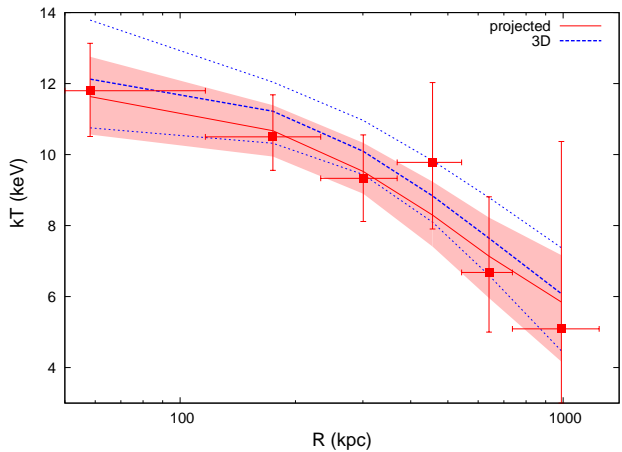}} \\
\caption[]{\label{f.ktprof}The projected temperature
profile of \jjj\ measured with \XMM\ is plotted along with the best fitting
 3D model and its projection.}
\end{center}
\end{figure}

The \XMM\ observation of \jjj\ is sufficiently deep to provide radial
profiles of its spectral properties. Spectra were extracted from annuli
centred on the X-ray centroid (with point sources excluded), and were fit as
before. The resulting projected temperature profile is plotted in
Fig. \ref{f.ktprof}. The data from the three EPIC camera were fit
simultaneously for these results, with the temperatures tied, the PN
absorbing column frozen at the Galactic value, and the MOS absobing column
at $4.1\times10^{20}\pcmsq$

To model the temperature profile of \jjj\ we follow the method of
\citet[][V06]{vik06a} of projecting a model of the three-dimensional (3D) 
temperature distribution along the line of sight, and fitting that to the observed projected
temperature profile. In doing this we use the algorithm of \citet{vik06b}
to predict the observed temperature for the combination of temperatures and
emissivities of gas along the line of sight in each bin. The emissivities
were determined from our analysis of the gas density profile (\textsection
\ref{s.gdens}), and the metal abundance of the gas was set at $0.3\Zsol$ at
all radii (varying the metal abundance had negligible effect on this
analysis, which is reasonable, as the emission is continuum dominated at
these high temperatures). The temperature profile model
of V06 was used to model the 3D temperature distribution:
\begin{eqnarray}\label{e.kt}
kT(r) & = & kT_0
\frac{(r/r_{cool})^{a_{cool}}+kT_{min}/kT_0}{(r/r_{cool})^{a_{cool}}+1} 
\frac{(r/r_t)^{-a}}{(1+(r/r_t)^b)^{c/b}}.
\end{eqnarray}
This model was simplified slightly to reduce the number of free parameters
by fixing $a=0$ and $b=c/0.45$, in line with the average profile found by
V06 in a sample of relaxed nearby clusters. The best fitting 3D model had $kT_0=22.0\keV$, $r_{cool}=47.4\kpc$,
$a_{cool}=0.0$, $kT_{min}=2.7\keV$, $r_t=329\kpc$, and $c=0.72$, and the
model and its projection are plotted in Fig. \ref{f.ktprof}. This model has
as many degrees of freedom as datapoints in the profile, and has
degeneracies between parameters, so the uncertainties on all dependent
quantities were computed by a Monte-Carlo method. A simpler model
\begin{eqnarray}\label{e.bkt}
kT(r) & = & kT_0 (1+r/r_t)^{-\alpha}
\end{eqnarray}
was also fit to the data, giving
$kT_0=13.3\keV$, $r_t=1.0\Mpc$ and $\alpha=1.1$. While the V06 model
was used for all results quoted subsequently, this simpler model gave
consistent results and uncertainties throughout.

This method of modelling the temperature profile does not take the effect
of the \XMM\ PSF into account. As the surface brightness profile of \jjj\
is not sharply peaked and the temperature profile bins are all $\ge15\arcs$
(the approximate FWHM of the PSF), the mixing effect of the PSF on the
temperature profile is not expected to be large. This was verified by
fitting the spectra from all radial bins simultaneously with absorbed APEC
models, using the {\it xmmpsf} model in XSPEC to model the redistribution
of photons by the
\XMM\ PSF. The resulting PSF-corrected temperature profile was consistent with
the original profile, with the noise fluctuations amplified and large
uncertainties. The original uncorrected temperature profile was used for
all further analysis.

\begin{figure}
\begin{center}
\scalebox{0.35}{\includegraphics*[angle=270]{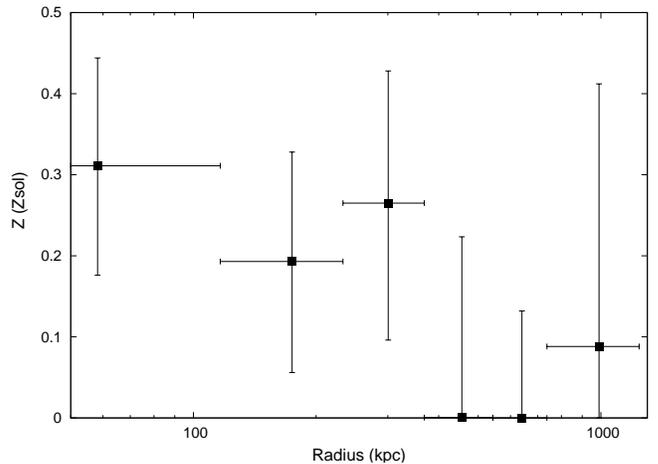}} \\
\caption[]{\label{f.Zprof}Projected metal abundance profile of \jjj. The y errorbars show the $68\%$ confidence intervals.}
\end{center}
\end{figure}

The projected metal abundance profile of \jjj\ is plotted in Fig
\ref{f.Zprof}. The statistical uncertainties on the metal abundance
measurements are large due to the high ionisation state of the plasma. The
data suggest a radially decreasing abundance profile but are also
consistent with a flat abundance profile.

\subsection{Temperature map}
The two dimensional projected temperature distribution in \jjj\ was
investigated by constructing a temperature map following the methods
described in \citet[][see also \citet{osu05}]{mau06b}. In brief, spectra
were extracted from many overlapping circular regions in the cluster, the
radii of which were chosen to contain at least $1000$ net photons. The best
fitting temperature for the spectrum from each region was assigned to the
pixel in the temperature map corresponding to the centre of that
region. The result is a smooth image of the projected temperatures, in
which nearby pixels are not independent. The spectra were fit as before,
but the metal abundance was fixed at $0.3\Zsol$.

\begin{figure*}
\begin{center}
\plotone{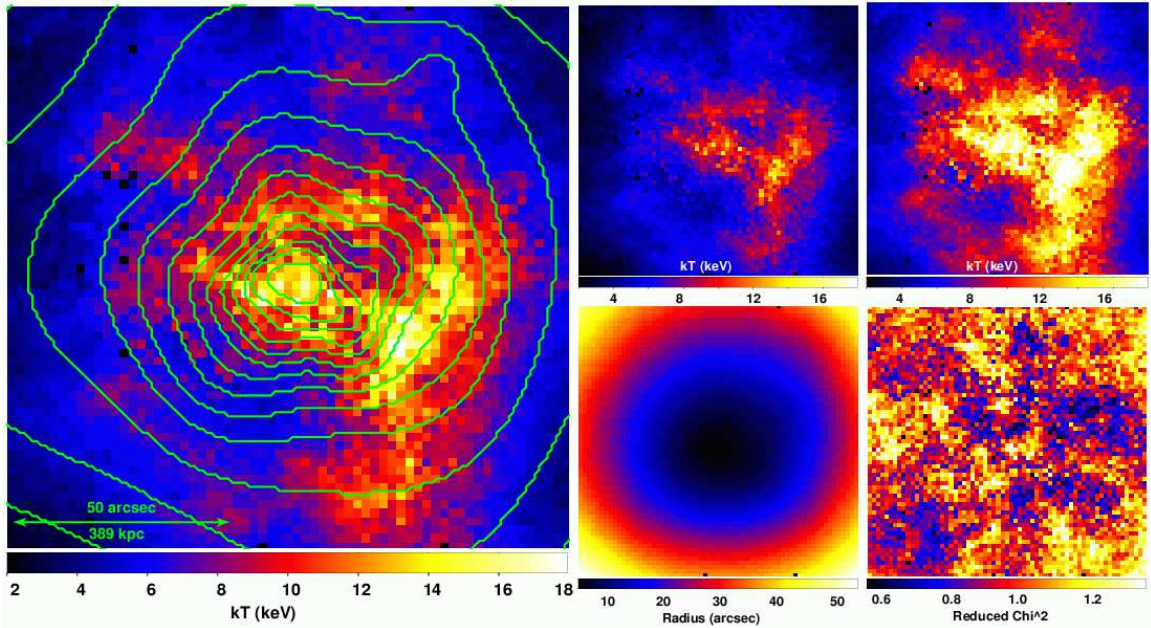}
\caption{\label{f.ktmap}{\it Left:} Projected temperature map of
\jjj\ with \Chandra\ contours of X-ray surface brightness overlaid. {\it
Right:} The upper panels show the $1\sigma$ upper
(left) and lower (right) limits on the temperature values. The lower panels
show the radii used for
spectral extraction and the reduced $\chisq$ of the spectral fits.} 
\end{center}
\end{figure*}

\begin{figure}
\begin{center}
\scalebox{1}{\plotone{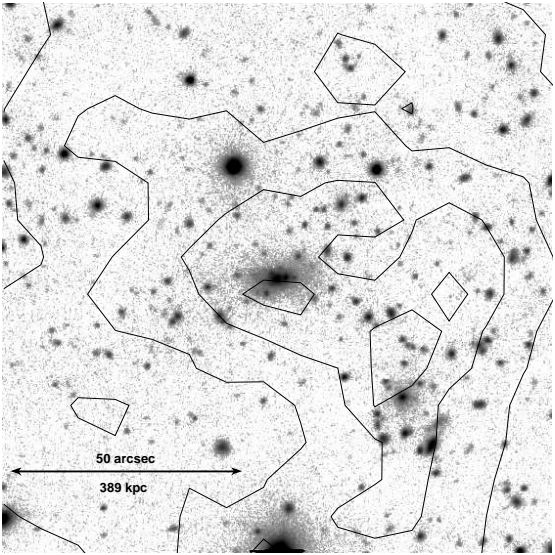}}
\caption{\label{f.ktcon}Contours of the projected X-ray temperature are
overlaid on the optical image of \jjj. Contours are set at 6, 9, 12, and $15\keV$.} 
\end{center}
\end{figure}

The resulting projected temperature map is shown in Fig. \ref{f.ktmap}. The
temperature structure shows asymmetry that is not apparent in the surface
brightness distribution. There is significantly hotter emission to the west
and south west of the cluster core than the surrounding gas at the same
radius. Fig. \ref{f.ktcon} shows contours of the X-ray temperature overlaid
on the optical image. The region of hotter emission appears associated with
an overdensity of galaxies (confirmed cluster members) to the south west of
the core, suggesting that a merger event may be responsible for the
temperature asymmetry.

A new temperature profile was produced, with the sector containing these
hotter regions excluded, and the temperatures obtained were not significantly
different from those measured with the sector included. This indicates that
this hot region does not significantly bias our temperature profile
measurement. However, if it is related to merger activity in the cluster,
this will undermine the assumption of hydrostatic equilibrium used in
the following mass analysis of \jjj.

\section{Mass Analysis}

\subsection{Gas density profile}\label{s.gdens}
The X-ray emissivity of the intra-cluster medium depends strongly on the
gas density and only weakly on its temperature. This means that the
observed projected emissivity profile can be used to accurately measure the
gas density profile. For each annular bin in the surface brightness
profile, the observed net count rate was corrected for area lost to chip
gaps, bad pixels and excluded regions, and converted to an integrated
emission measure. For each bin, the conversion factor was calculated
assuming an absorbed \MEKAL\ \citep{kaa93} plasma model folded through an
ARF generated for that region, and an on-axis RMF. The absorption of the
spectral model was set at the galactic value and the metal abundance was
set at $0.3$. The temperature of the spectral model in each bin was given
by the best-fitting 3D temperature model. We note that the dependence of
the conversion from count rate to emission measure on the assumed
temperature is weak; this analysis was repeated assuming the cluster was
isothermal for the purpose of deriving the emission measures, and the
results were not significantly changed.

\begin{figure}
\begin{center}
\scalebox{0.35}{\includegraphics*[angle=270]{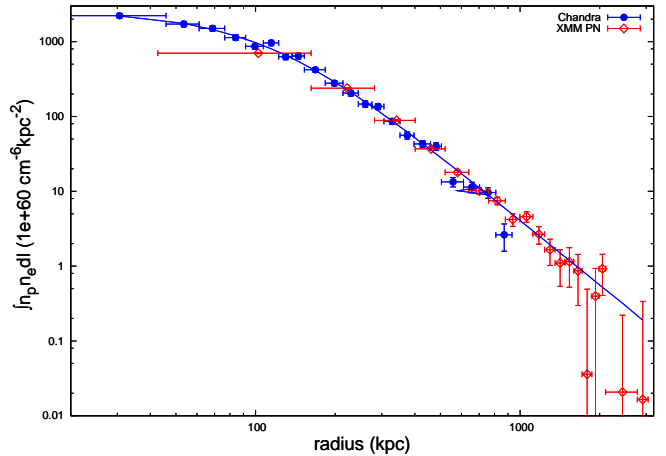}} \\
\caption[]{\label{f.em}The projected emission measure profile of \jjj\ as
measured with \Chandra\ and \XMM\ is plotted along with the best fitting model.}
\end{center}
\end{figure}

Projected emission measure profiles were derived from the ACIS and PN data
in this way, and are plotted in Fig. \ref{f.em}. The radial bins were set
to a minimum width of $15\arcs$ for the PN profile, to minimise the effect
of the \XMM\ PSF, and the profiles agree very well. The MOS data were omitted
here for simplicity, as the PN data alone were sufficient to extend the
profile beyond the range of the temperature profile, which is the limiting
factor in the mass analysis.

To model the emissivity profile, we used
the modified $\beta$-profile used by V06 to describe the
emission measure profile of the gas in nearby relaxed clusters;
\begin{eqnarray}\label{e.emis}
n_pn_e & = & n_0^2 \frac{(r/r_c)^{-\alpha}}{(1+r^2/r_c^2)^{3\beta-\alpha/2}}(1+r^\gamma/r_s^\gamma)^{-\epsilon/\gamma}.
\end{eqnarray}
This model can describe a steep core component and any steepening of the
data beyond the standard $\beta$-profile at large radii.
A additive core component has been removed from the V06 model here as it
was not required by the data.  This emission measure model was then
projected along the line of sight and fit to the observed projected
emission measure profile. The model was fit to the \Chandra\ and \XMM\ PN
data simultaneously, with the PN data within $500\kpc$ excluded to minimise
PSF effects. The best fitting model is plotted in Fig. \ref{f.em}. The gas
density is then related to the emission measure by
$\rho_g=1.252m_p(n_pn_e)^{1/2}$, assuming a cosmic plasma with helium
abundances given by \citet{and89}. For comparison with other work, the
profile was also fit with a standard $\beta$-model, and the best-fitting
parameters were $r_c=105\pm5\kpc$ and $\beta=0.64\pm0.01$.

\subsection{Mass profiles}
\begin{figure}
\begin{center}
\scalebox{0.35}{\includegraphics*[angle=270]{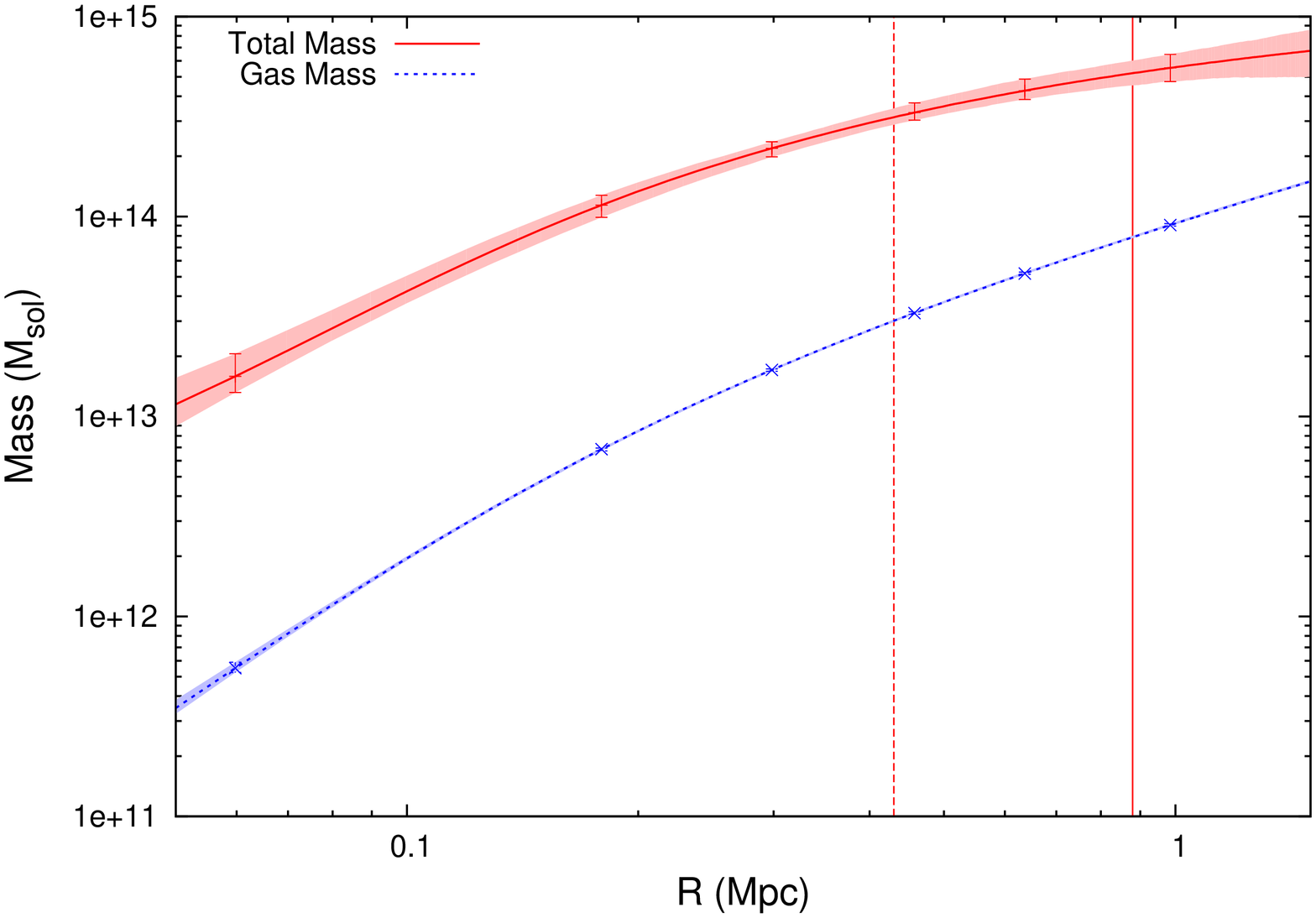}} \\
\caption[]{\label{f.mprof}Total (upper line) and gas mass (lower line)
profiles of \jjj. The data points indicate the midpoints of the temperature
bins and the shaded regions and errorbars show the $1\sigma$ confidence
intervals from the Monte-Carlo randomisations of the data. The vertical
lines indicates the radii of \rn{2500}\ and \rf.}
\end{center}
\end{figure}

The 3D models of the gas density and temperature
distributions enable the total gravitating mass of \jjj\ to be derived
under the assumptions of spherical symmetry and hydrostatic
equilibrium. Fig. \ref{f.mprof} shows the total and gas mass profiles of
\jjj\ with the shaded regions indicating the $1\sigma$ uncertainties. The
error analysis was performed by producing 1000 randomisations of the
observed temperature and emission measure profiles based on their
measurement uncertainties. The full analysis was repeated for each
randomisation, and the $\pm34$ percentiles of the distribution about the
value determined from the non-randomised data gave the $\pm1\sigma$
uncertainties on each parameter.

\begin{figure}
\begin{center}
\scalebox{0.35}{\includegraphics*[angle=270]{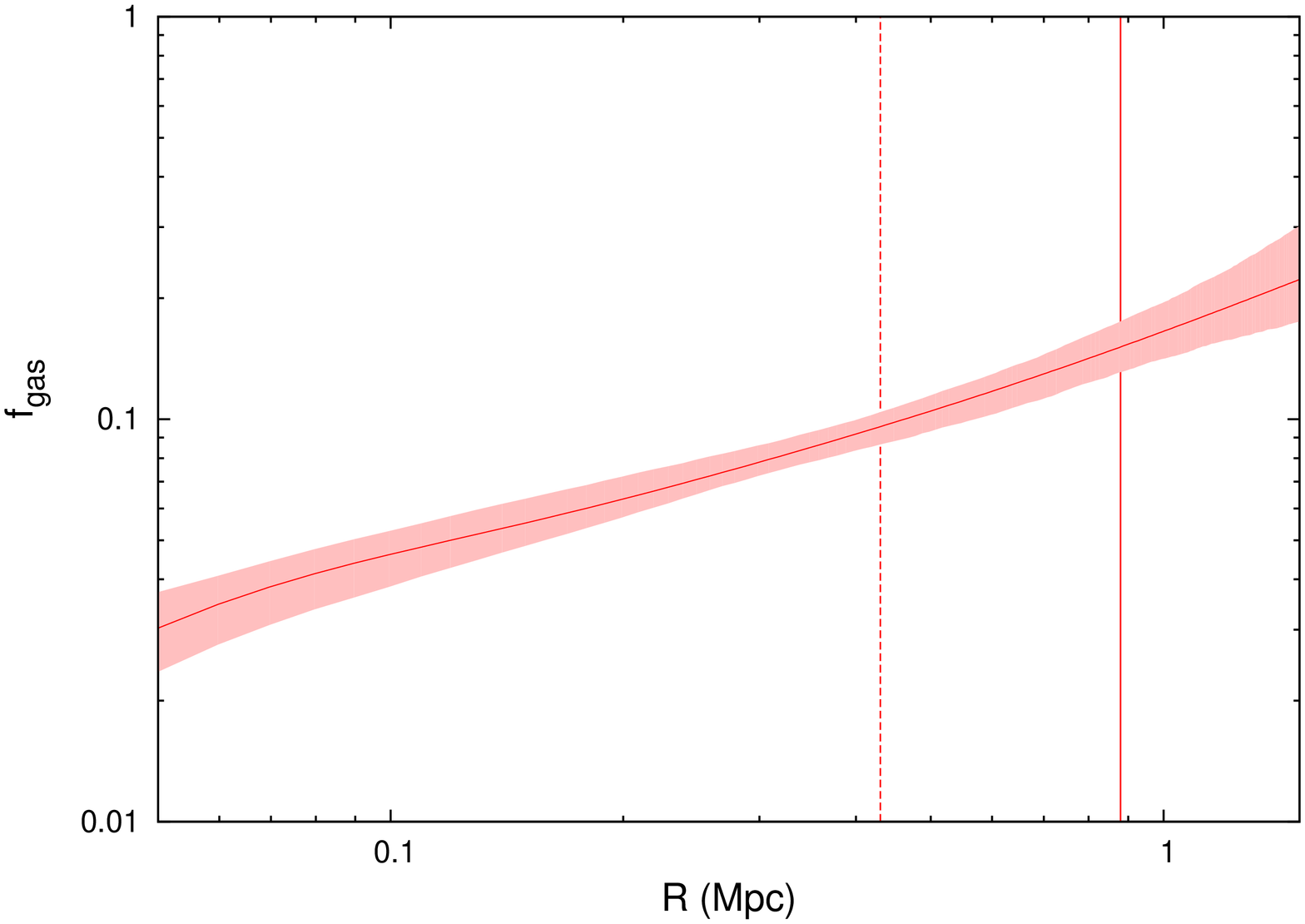}} \\
\caption[]{\label{f.fgas}The gas mass fraction of \jjj\ is plotted as a
function of radius. The shaded region shows the $1\sigma$ confidence
interval and the vertical lines mark the radii of \rn{2500}\ and \rf.}
\end{center}
\end{figure}

The enclosed gas mass fraction (\fgas) profile of \jjj\ was also computed,
and the resulting profile is plotted in Fig. \ref{f.fgas}. The profile
increases out to \rf, which is consistent with the profiles measured by V06
for local clusters, but contrasts with the findings of \citet{all04} that
\fgas\ profiles tend to a universal value at around \rn{2500}\ (although
with some variation between clusters). The latter work used \Chandra\
observations of sample of clusters including \jjj; the value of \fgas\ we
find for \jjj\ at \rn{2500} is consistent with that measured by \citet{all04}.

\section{Entropy and cooling time profiles}
\begin{figure}
\begin{center}
\scalebox{0.35}{\includegraphics*[angle=270]{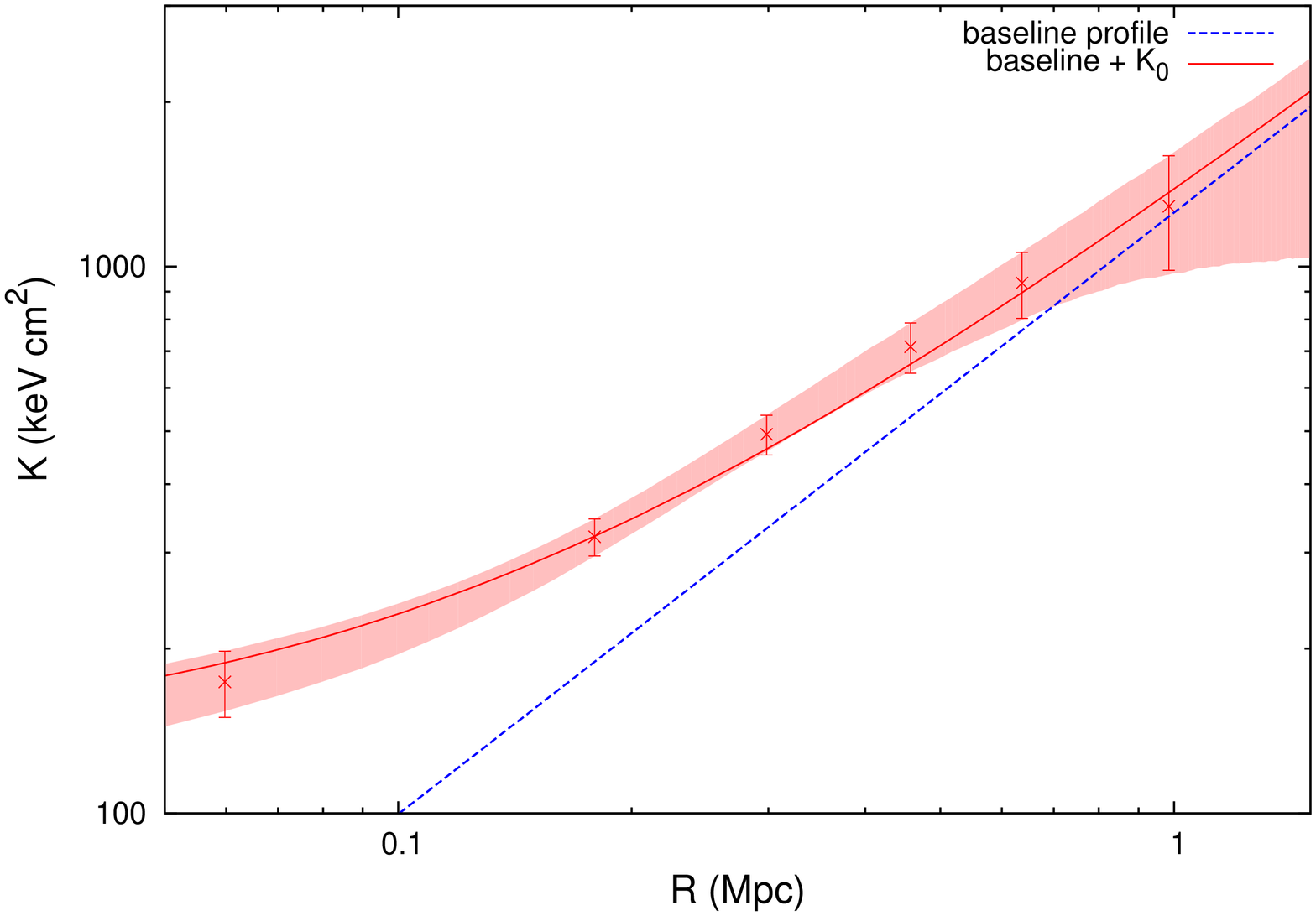}} \\
\caption[]{\label{f.kprof}The gas entropy of \jjj\ is plotted as a function
of radius with the shaded region indicating the $1\sigma$ confidence
intervals and the data points the measured values at the midpoint of the
temperature bins. The dotted line shows the baseline entropy profile of
\jjj\ due to gravitational effects alone \citep{voi05c} and the solid line
is the baseline profile with an additional constant entropy level (see text).}
\end{center}
\end{figure}

The entropy ($K$) of the intra-cluster gas can give useful insight into its
energetics \citep[\egc][]{pon99,voi05b}. The entropy profile of
\jjj\ was derived from the 3D temperature and gas density models according to the definition $K=kT n_e^{-2/3}$ and is
plotted in Fig. \ref{f.kprof}. In order to model this entropy profile, we
started with the baseline entropy profile due to gravitational effects alone,
derived from numerical simulations by \citet{voi05c};
$K(r)=1.32K_{200}(r/R_{200})^{1.1}$. $K_{200}$ was estimated following the
method suggested by \citet{voi05c}, with
\begin{eqnarray}\label{e.k200}
K_{200} & = & 362kT_{200}E(z)^{-4/3},
\end{eqnarray}
where $E(z)$ gives the redshift-variation of the Hubble parameter and
$kT_{200}$ was estimated from the mass within \rt\ via
$kT_{200}=GM_{200}\mu m_p/2R_{200}$. For \jjj\ $kT_{200}$ was found to be
$6.6\keV$ by extrapolating the mass profile model to \rt, and hence
$K_{200}=1226\ent$. The baseline entropy profile is plotted in
Fig. \ref{f.kprof} and is clearly not a good description of the data. The
data suggest a large entropy excess above the gravitational baseline in the
central regions of
\jjj. To model this, a constant entropy level ($K_0$) was added to the baseline
model to give
\begin{eqnarray}\label{e.kprof}
K(r)=K_0 + 1.32K_{200}(r/R_{200})^{1.1}.
\end{eqnarray}

This model was fit to the entropy values measured at the midpoints of the 6
temperature bins of \jjj. The best fitting model had $K_0=132\pm24\ent$ and is plotted in
Fig. \ref{f.kprof}. The model is a good description of the data
($\chisq/\nu=1.7/5$) and a significant improvement on the baseline model
alone. \citet{don06} used a similar model to
fit entropy profiles in nearby cooling flow clusters, and the central entropy
level $K_0$ is significantly higher in \jjj\ than the values of $\sim10\ent$
found by \citet{don06}. This is consistent with the high central
temperature and lack of cool core in \jjj, and is likely a result of the
suspected merger activity in the system. 

\begin{figure}
\begin{center}
\scalebox{0.35}{\includegraphics*[angle=270]{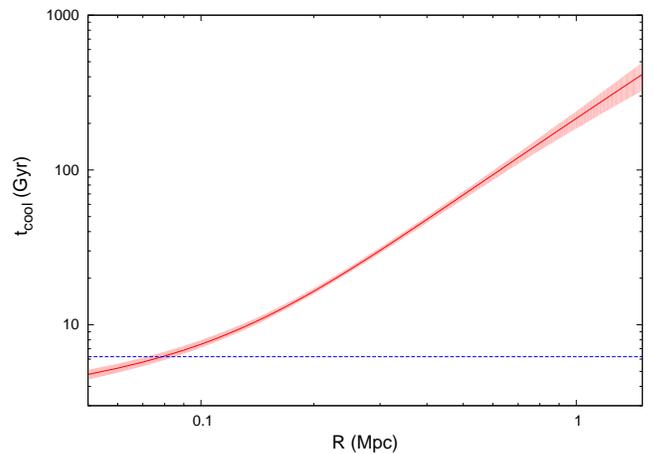}} \\
\caption[]{\label{f.tcprof}The cooling time of the gas in \jjj\ is plotted
as a function of radius. The shaded region shows the $1\sigma$ confidence
interval and the horizontal line marks the age of the universe
at $z=0.89$.}
\end{center}
\end{figure}

The 3D gas density and temperature profiles enable the cooling time profile
of \jjj\ to be computed, using the relation
\begin{eqnarray}
t_{cool} & = &
8.5\times10^{10}\yr\left(\frac{n_p}{10^{-3}\pcc}\right)\left(\frac{T}{10^{8}\K}\right)
\end{eqnarray}
\citep{sar86}. The resulting profile is plotted in Fig. \ref{f.tcprof}
along with the Hubble time at $z=0.89$, and shows that the gas in the
central regions of \jjj\ would take $\sim6\Gyr$ to radiate its thermal
energy away. This combined with possibility of a recent merger event
(implied by the asymmetry in the
temperature map and galaxy distribution) make the absence of cool gas in
the core of this system unsurprising.
As pointed out by \citet{cag01} there is a faint point-like radio source at
the centre of \jjj, consistent with a low-luminosity radio-loud AGN. This is also a candidate for energy input into the
cluster core at some point in the past.

\section{Discussion}
The deep \XMM\ observation of \jjj\ has enabled a uniquely detailed mass
analysis of a $z=0.89$ galaxy cluster. This allows comparison with
theoretical density profiles for the first time at such high redshift. A
profile was constructed using the densities calculated at the
midpoint of each temperature profile bin, with uncertainties derived from
the Monte-Carlo realisations. These data points were fit by a
\citet[][NFW]{nav96,nav97} profile:
\begin{eqnarray}\label{e.nfw}
\frac{\rho(r)}{\rho_{c}} & = & \frac{\delta_c}{(r/r_s)(1+r/r_s)^2}
\end{eqnarray}
where the normalisation $\delta_c$ can be expressed in terms of the halo
concentration $c_\Delta$, and the overdensity factor of interest $\Delta$:
\begin{eqnarray}\label{e.nfw2}
\delta_c & = & \frac{\Delta}{3}\frac{c^3}{ln(1+c)-c/(1+c)}.
\end{eqnarray}
The scale radius $r_s$ is then related to the overdensity radius $R_\Delta$
by the concentration, $c_\Delta=R_\Delta/r_s$. This model has been used
successfully to describe the density profiles of both simulated and
observed clusters of a range of masses. 

\begin{figure}
\begin{center}
\scalebox{0.35}{\includegraphics*[angle=270]{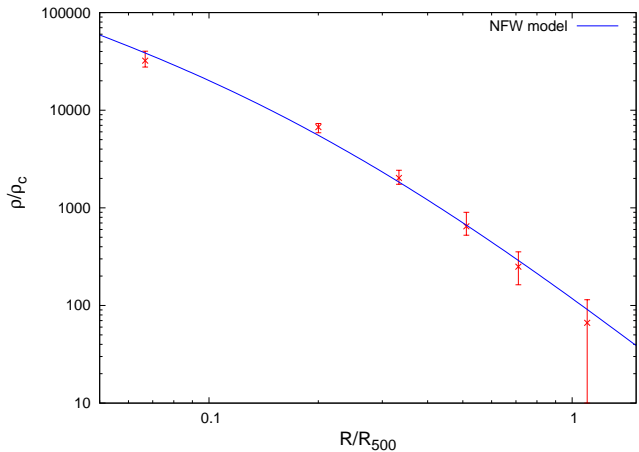}} \\
\caption[]{\label{f.nfw}The total and gas density profiles of \jjj\ are
plotted, scaled by the critical density at $z=0.89$ and \rf. The solid line
is the best fitting NFW profile to the data.}
\end{center}
\end{figure}

The model was fit to the density profile of \jjj\ using $\Delta=500$, and
the best-fitting parameters were $c_{500}=5.3^{+1.2}_{-1.0}$ and
$r_s=0.18\pm0.04\Mpc$ ($\chisq/\nu=4.2/4$). Using $\Delta=200$ scales the
concentration to $c_{200}=7.9^{+1.7}_{-1.4}$. The data and best fitting
model are plotted in Fig. \ref{f.nfw}. The measured concentration of \jjj\
is significantly higher than that found for local clusters of a similar
mass by V06. Furthermore, simulations predict that $c_\Delta$ should
{\it decrease} with increasing redshift (for a fixed mass)
\citep[\egc][]{wec02,bul01}. \citet{wec02} also found that concentration
increases with the time between the epoch at which a cluster is observed (\zo)
and that at which it formed (\zf); $c_\Delta\propto(1+\zo)/(1+\zf)$. With
$\zf\sim2.6$, the concentration of \jjj\ would be in line with these
predictions. However, given the suspected merger activity in \jjj,
$\zo\approx\zf$ seems more likely, with the high concentration resulting
from the merger activity.

\subsection{The global properties of \jjj\ and the scaling relations}
To enable comparisons with other works, we compute the global
properties of \jjj\ within different radii. These are summarised in Table
\ref{t.summ}. The temperatures and metal abundances were obtained from
single temperature fits to the spectra extracted within each radius using
the combined \XMM\ data. These represent some of the most precise such
measurements for a high-redshift system. The metal abundance is slightly
lower than the values of $Z\approx0.3\Zsol$ observed in local clusters
\citep[\egc][]{fuk98a}, but is consistent with the declining abundance with
redshift predicted by \citet{ett05} based on modelled supernovae rates.

We computed the bolometric
luminosity of the cluster by scaling the value measured within \rf\ out to
$1.43\Mpc$ \citep[for consistency with][]{mar98a} and out to infinity using
the measured surface brightness profile. We find
$L_X(r<1.43\Mpc)=4.57\pm0.11\ergps$ and
$L_X(r<\infty)=5.12\pm0.12\ergps$, including uncertainties on the
temperature and metal abundance in those on the luminosity. These luminosities are consistent with those
found by \citet{mau06a} based on a shorter \XMM\ observation of \jjj.

\begin{table*}[t]
\centering 
\begin{tabular}{cccccccc} \hline 
$\Delta$ & $r_\Delta (Mpc)$ & $kT (keV)$ & $Z (Z_\odot)$ & $M_{gas}
(10^{13}\Msol)$ & $M_{tot} (10^{14}\Msol)$ & $f_{gas}$ & $\Yx (10^{14}\Msol\keV)$ \\ \hline 

$500$ &$0.88\pm0.05$ & $10.4\pm0.6$ & $0.16\pm0.05$ & $7.8^{+0.7}_{-0.5}$ &
$5.2^{+1.0}_{-0.8}$ & $0.15^{+0.02}_{-0.01}$ & $8.1^{+0.9}_{-0.7}$ \\ 


$2500$ & $0.43\pm0.02$ & $10.7\pm0.5$ & $0.16\pm0.05$ & $3.0\pm0.2$ &
$3.2^{+0.5}_{-0.4}$ & $0.10\pm0.01$ & $3.2\pm0.3$ \\ 


\end{tabular}
\caption{\label{t.summ}Summary of the global properties of \jjj\ measured
within different overdensity radii. The uncertainties include the
statistical uncertainties and the effect of a $50\%$ increase on the
temperature errors to account for calibration uncertainties.
}
\end{table*} 

\begin{figure}
\begin{center}
\scalebox{0.35}{\includegraphics*[angle=270]{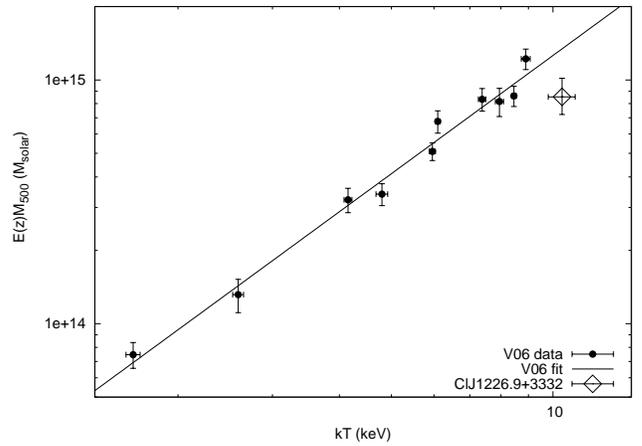}} \\
\caption[]{\label{f.mt}\jjj\ is plotted on the V06 \MT\ relation. The
masses have been scaled by the expected self-similar evolution.}
\end{center}
\end{figure}

The measured properties of \jjj\ were then compared with local
luminosity-temperature (\LT) relation of \citet{mar98a}:
\begin{eqnarray}
\frac{L_X(r<1.43\Mpc)}{10^{44}\ergps} & = & 6.35 E(z)\left(\frac{kT}{6\keV}\right)^{2.64}.
\end{eqnarray}
Here we have converted the \LT\ relation to our cosmology, and included the
predicted self-similar evolution. The properties of \jjj\ were found to agree
very well with the predictions of this relation. \jjj\ was then compared to the local mass-temperature (\MT) relation
derived by V06
\begin{eqnarray}
\frac{M_{500}}{10^{14}\Msol} & = & 2.93 E(z)^{-1}\left(\frac{kT_{500}}{5\keV}\right)^{1.61}.
\end{eqnarray}
The \MT\ relation is again converted to our cosmology with the predicted
self similar evolution. The V06 data and \MT\ relation are plotted in
Fig. \ref{f.mt} along with the data point for \jjj, which falls
significantly below the V06 \MT\ relation. 

\begin{figure}
\begin{center}
\scalebox{0.35}{\includegraphics*[angle=270]{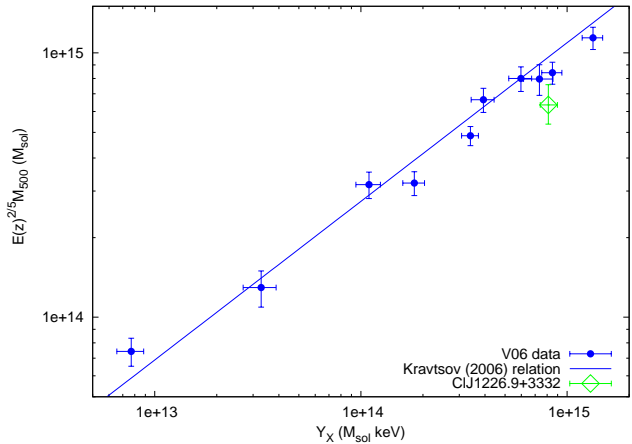}} \\
\caption[]{\label{f.ym}\jjj\ is plotted on the \citet{kra06a} \YM\ relation
along with the V06 data. The masses have been scaled by the expected
self-similar evolution.}
\end{center}
\end{figure}

Finally, \citet{kra06a} recently proposed a new mass proxy for clusters, \Yx,
obtained by multiplying the gas mass and temperature measured within \rf\
together. This quantity was found to scale with mass with smaller scatter than
other X-ray observables, and simulations showed that the scatter is less
than $8\%$ even for unrelaxed clusters, and that the scaling relation
evolves self-similarly. The high quality X-ray data for \jjj\ enable us
to test these predictions observationally for the first time at high
redshift. In Fig. \ref{f.ym} we reproduce the \YM\ plot from
\citet{kra06a} including the V06 data points. Following \citet{kra06a} we
parameterise the \YM\ relation as
\begin{eqnarray}
E(z)^{2/5}\frac{M_{500}}{10^{14}\Msol} & = & 1.58 \left(\frac{\Yx}{4\times10^{13}\Msol\keV}\right)^{3/5}.
\end{eqnarray}
Here, the normalisation found from the simulations of \cite{kra06a} has been
reduced by $15\%$ (in line with the recommendation in that work), to better
agree with the observed V06 cluster data. \jjj\ has a mass that is slightly
low compared to the local clusters (although it is within the scatter at
the $1\sigma$ level). This could be due to stronger than expected evolution
in the \YM\ relation, with $E(z)^{\sim4/5}$ resulting in good
agreement. A more likely explanation though, is that the offset from the relation
is caused by errors in one or both parameters. The simulations of
\citet{kra06a} show that the scatter in \YM\ is very small, even for merging
clusters. Taking the measured \Yx\ of \jjj\ as an indicator of the true
total mass then implies that the X-ray analysis underestimates the
true mass by $\sim30\%$. Such a difference would also improve the agreement
of \jjj\ with the \MT\ relation. 

Numerical simulations have shown that merging clusters tend to have lower
temperatures for a given mass than relaxed systems, as the increase in
thermal energy of the gas does not match the increase in mass due to the
merger \citep{mat01b,kra06a}. Thus the suspected merger event in \jjj\ could
be responsible for an underestimate of the total mass using X-ray
methods. The good agreement of \jjj\ with the \LT\ relation is not
inconsistent with this explanation, as mergers have been found to boost the
luminosity and temperature of clusters approximately simultaneously, moving
them along the \LT\ relation \citep{ran02,row04}.

\subsection{The effect of systematic temperature errors}
As discussed in \textsection \ref{s.cal} an artificially high absorbing
column was required when fitting the MOS spectra to produce agreement with
the PN data. It appears likely that the PN data are the more reliable in
this case, as the best-fitting PN absorbing column is close to the Galactic
value while the MOS is significantly higher. However, the full analysis was
also repeated with the MOS column fixed at the Galactic value, resulting in
systematically higher temperatures in the combined EPIC fit. \citet{vik05a}
also found cluster temperatures measured with \Chandra\ that were
systematically higher by $\sim10\%$ than those measured by \XMM. It is
thus instructive to evaluate the effect that systematic increase to the
individual temperatures would have on the inferred properties of \jjj.

Reassuringly, the total mass, gas mass, overdensity radii, and luminosity
of \jjj\ did not change significantly when the Galactic column was used for
the MOS data. These quantities are all robust to residual uncertainties in
the EPIC calibration. The only significant change was in the temperatures,
with the global temperature increasing to $11.8\keV$, with a corresponding
increase in $\Yx$. The agreement with the \LT, \MT, and \YM\ relations were
thus all worsened by the higher $kT$. The derived mass did increase
slightly in this test, but remained within our uncertainties because we
included a large systematic contribution in our error budget and because
the shape of the temperature profile did not change significantly in this
experiment, even though its normalisation did. An isothermal mass estimate
would have been much more sensitive to the increase in $kT$.

\section{Summary and Conclusions}
High quality \XMM\ and \Chandra\ data of \jjj\ have enabled the most
precise X-ray mass analysis of any such high-redshift system. The
temperature profile of the system showed no sign of central cooling, with a
hot core and a radially declining profile. The cooling time of the gas only
drops marginally below the Hubble time in the centre of the cluster. A
temperature map showed asymmetry in the temperature distribution with a hot
region that is apparently associated with a subclump of galaxies, but is
not visible in the X-ray surface brightness. This is likely to be result of
a merger event in the cluster, but does not appear to significantly affect
the overall temperature profile.

The entropy profile in \jjj\ was not well described by the gravitational
baseline profile of \citet{voi05c} and required an additional constant
entropy level. The \fgas\ profile increases with radii
out to \rf, and is consistent with the profiles observed in local
clusters. The global properties of \jjj\ were
compared with the local scaling relations and found to be consistent with
the \LT\ relation, but to fall below the \MT\ and \YM\ relations. As \Yx\
is insensitive to mergers, this implies that the total mass is
underestimated by $\sim30\%$ in the X-ray analysis, probably due to
departures of the gas from hydrostatic and virial equilibrium. The high NFW
concentration parameter found for \jjj\ is likely also due to the merger
activity. Given the high luminosity of \jjj, it is likely to be the only
distant cluster for which such rich X-ray data are obtained with the
current generation of satellites.

\acknowledgments
We thank Alexey Vikhlinin for providing some of the software used in this work
and Ewan O'Sullivan for useful discussions of the \XMM\ background
treatment. BJM is supported by NASA through Chandra Postdoctoral
Fellowship Award Number PF4-50034 issued by the Chandra X-ray Observatory
Center, which is operated by the Smithsonian Astrophysical Observatory for
and on behalf of NASA under contract NAS8-03060.

\appendix
\section{A significance contouring method}\label{s.con}
Here we describe the procedure used to define contour levels based on the
significance of the emission enclosed between them. For this method to
succeed, the image used to define the contour regions needs to be
sufficiently smoothed that smooth contiguous regions can be defined even
where the number of counts are low. An adaptive smoothing algorithm is most
suited for this purpose as it uses a large smoothing kernel in regions of
low signal to noise, but maintains high spatial resolution where the signal
to noise ratio is high. Two adaptive smoothing algorithms were tested; {\it
asmooth} \citep[][upon which the CIAO task {\it csmooth}\footnote{Note
however that the CIAO {\it csmooth} is inferior to the \citet{ebe06a} {\it
asmooth} algorithm. This is addressed by \citet{ebe06a} and some
shortcomings of {\it csmooth} are pointed out by \citet{die05}.} is
based]{ebe06a}, and the \XMM\ Science Analysis Software (SAS) smoothing tool
(this tool is also called {\it asmooth}, but for clarity we refer to it
herein as {\it xmmsmooth}). We found that for this particular purpose, the
{\it asmooth} was more suitable than {\it xmmsmooth} as {\it xmmsooth}
created artifacts around point sources and pixelation in the cluster cores
(even with a high smoothing significance threshold) that were incompatible
with our contouring algorithm. We note that both {\it asmooth} and {\it
xmmsooth} are less sensitive to regions of depressed emission than regions
of excess emission.

To set the lowest contour level, the flux was measured in the background
image for that cluster (after it had been normalised to match the local
background level) in a circle of radius $6\arcsec$ at the cluster
centre. The lowest contour level was then set at a flux of $3\sigma$ above
this background level. The flux interval between this level ($f_b$) and the
peak of the cluster flux was divided into 500 logarithmically spaced
intervals ($f_i$; $i=1-500$). The next contour level was set at $f_1$, and
spatial regions bounded by the contours at $f_b$ and $f_1$ were defined on
a smoothed image of the X-ray emission. For each distinct region, the
significance of the emission contained in that region above the background
was measured using the counts in the original unsmoothed source and
background images. Regions that contained emission detected at $>3\sigma$
significance were kept, and the others were rejected. 

The process was then repeated with an upper flux limit of $f_2$. Where
regions had passed the previous step, new ``child'' regions were defined
bounded by the contours at $f_1$ to $f_2$, and were by definition fully
enclosed by the passing parent region from the previous step. Where regions
failed the previous step, new regions bounded by contour levels $f_b$ and
$f_2$ were defined. This process was then repeated for all of the flux
levels $f_i$, with the significance of the emission enclosed by each region
measured with respect to its local background. The local background for all
child regions was measured in their respective parent region. For the
lowest level regions bounded by contours at $f_b$ and $f_i$, the blank-sky
background image was used to measure the background level. 

\begin{figure}
\begin{center}
\plotone{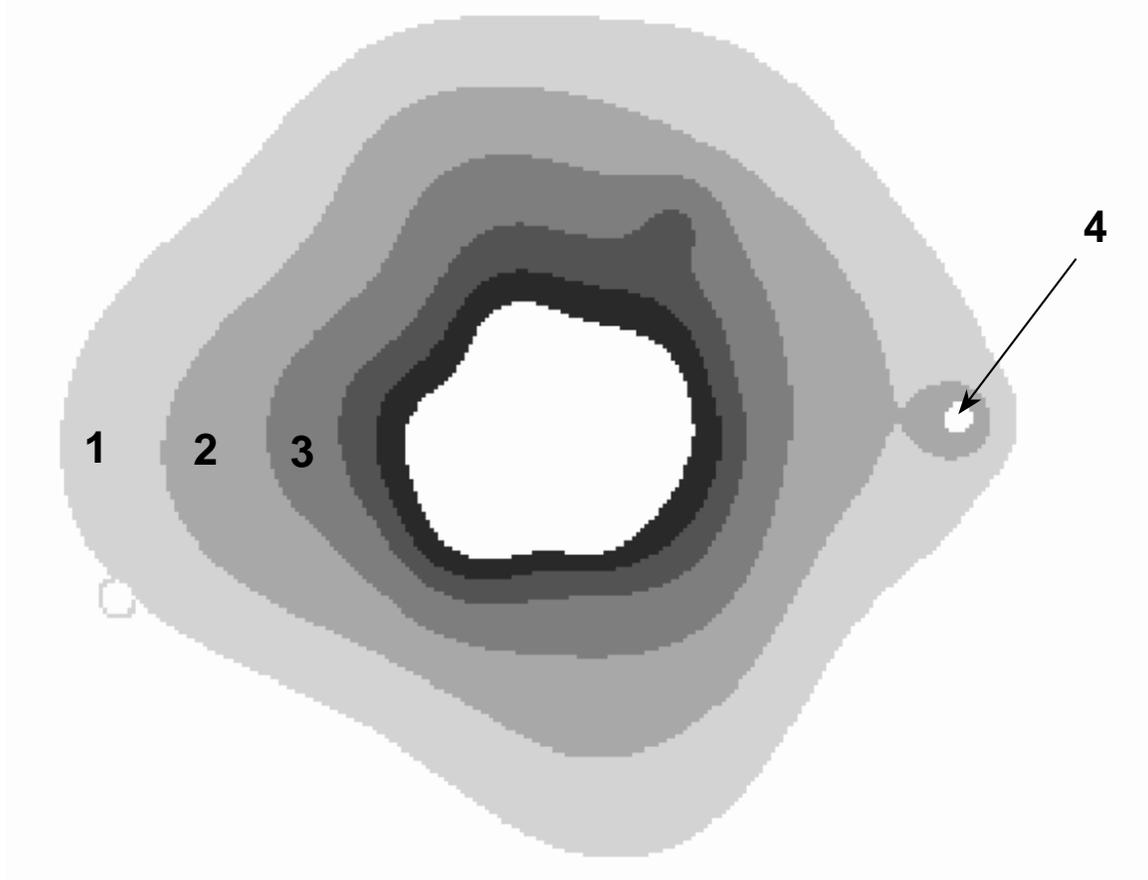}
\caption{\label{f.reg}Some of the regions used to define the contour levels
in the \Chandra\ image of \jjj.} 
\end{center}
\end{figure}

An image of some of the regions used to define the contour levels in the
\Chandra\ image of \jjj\ is shown in Fig. \ref{f.reg} for
illustration. Region 1 is bounded by contour levels at $f_b$ and $f_1$, and
contains emission detected at $>3\sigma$ above the emission in region 1 in
the background image. Region 2 uses the emission in region 1 as the
background level above which to measure its significance. Regions 3 and 4
use region 2 as a background, but their significance is measured
independently. Note that this contouring scheme means that contour levels
cannot be interpreted in the same way as in standard contour plots; separate
contours that are the same number of contours above the background are
unlikely to correspond to the same flux level in the smoothed image.

When applying this method to a large number of \Chandra\ observations of
clusters (Maughan \etal 2006 in prep.) it was found that an artificial
``banding''could result. This is a repeated sequence of two tightly grouped
contours followed by two widely spaced contours. The reason for this is
that the statistical uncertainties on the blank-sky background counts in a
region are generally very small, so a narrow contour band is sufficient for
a $3\sigma$ detection. When that narrow band is used for a background for
its child region, the uncertainties on the background level are relatively
large due to the small number of counts, so a wide contour band is required
for a $3\sigma$ detection. The next band is then narrow because of the
relatively small background uncertainties in the wide parent region, and
the process repeats. In order to prevent this, an additional criterion was
imposed to regulate the process. In order for a region to pass, it was
required that the contained emission be detected at $>3\sigma$ {\it and}
that the statistical uncertainty on the enclosed flux be smaller than some
threshold. A threshold of $5\%$ was used for the majority of the clusters,
but a higher (lower) threshold was required for the faintest (brightest)
clusters. This regulation prevents the banding problem and means that,
particularly in the brightest clusters, the significance of some regions is
somewhat larger than the $3\sigma$ threshold.

\clearpage

\bibliographystyle{mn2e}
\bibliography{clusters}

\begin{thebibliography}{}

\bibitem[\protect\citeauthoryear{{Allen}, {Schmidt}, {Ebeling}, {Fabian} \&
  {van Speybroeck}}{{Allen} et~al.}{2004}]{all04}
{Allen} S.~W.,  {Schmidt} R.~W.,  {Ebeling} H.,  {Fabian} A.~C.,    {van
  Speybroeck} L.,  2004, \mnras, 353, 457

\bibitem[\protect\citeauthoryear{{Anders} \& {Grevesse}}{{Anders} \&
  {Grevesse}}{1989}]{and89}
{Anders} E.,  {Grevesse} N.,  1989, \gca, 53, 197

\bibitem[\protect\citeauthoryear{{Arnaud}, {Majerowicz}, {Lumb}, {Neumann},
  {Aghanim}, {Blanchard}, {Boer}, {Burke}, {Collins}, {Giard}, {Nevalainen},
  {Nichol}, {Romer} \& {Sadat}}{{Arnaud} et~al.}{2002}]{arn02b}
{Arnaud} M.,  {Majerowicz} S.,  {Lumb} D.,  {Neumann} D.~M.,  {Aghanim} N.,
  {Blanchard} A.,  {Boer} M.,  {Burke} D.~J.,  {Collins} C.~A.,  {Giard} M.,
  {Nevalainen} J.,  {Nichol} R.~C.,  {Romer} A.~K.,    {Sadat} R.,  2002, \aap,
  390, 27

\bibitem[\protect\citeauthoryear{{Bullock}, {Kolatt}, {Sigad}, {Somerville},
  {Kravtsov}, {Klypin}, {Primack} \& {Dekel}}{{Bullock} et~al.}{2001}]{bul01}
{Bullock} J.~S.,  {Kolatt} T.~S.,  {Sigad} Y.,  {Somerville} R.~S.,  {Kravtsov}
  A.~V.,  {Klypin} A.~A.,  {Primack} J.~R.,    {Dekel} A.,  2001, \mnras, 321,
  559

\bibitem[\protect\citeauthoryear{{Cagnoni}, {Elvis}, {Kim}, {Mazzotta}, {Huang}
  \& {Celotti}}{{Cagnoni} et~al.}{2001}]{cag01}
{Cagnoni} I.,  {Elvis} M.,  {Kim} D.-W.,  {Mazzotta} P.,  {Huang} J.-S.,
  {Celotti} A.,  2001, \apj, 560, 86

\bibitem[\protect\citeauthoryear{Dickey \& Lockman}{Dickey \&
  Lockman}{1990}]{dic90}
Dickey J.~M.,  Lockman F.~J.,  1990, \araa, 28, 215

\bibitem[\protect\citeauthoryear{{Diehl} \& {Statler}}{{Diehl} \&
  {Statler}}{2005}]{die05}
{Diehl} S.,  {Statler} T.~S.,  2005, ArXiv Astrophysics e-prints

\bibitem[\protect\citeauthoryear{{Donahue}, {Gaskin}, {Patel}, {Joy}, {Clowe}
  \& {Hughes}}{{Donahue} et~al.}{2003}]{don03}
{Donahue} M.,  {Gaskin} J.~A.,  {Patel} S.~K.,  {Joy} M.,  {Clowe} D.,
  {Hughes} J.~P.,  2003, \apj, 598, 190

\bibitem[\protect\citeauthoryear{{Donahue}, {Horner}, {Cavagnolo} \&
  {Voit}}{{Donahue} et~al.}{2006}]{don06}
{Donahue} M.,  {Horner} D.~J.,  {Cavagnolo} K.~W.,    {Voit} G.~M.,  2006,
  \apj, 643, 730

\bibitem[\protect\citeauthoryear{Ebeling, Jones, Fairley, Perlman, Scharf \&
  Horner}{Ebeling et~al.}{2001}]{ebe01a}
Ebeling H.,  Jones L.~R.,  Fairley B.~W.,  Perlman E.,  Scharf C.,    Horner
  D.,  2001, \apj, 548, L23

\bibitem[\protect\citeauthoryear{{Ebeling}, {White} \& {Rangarajan}}{{Ebeling}
  et~al.}{2006}]{ebe06a}
{Ebeling} H.,  {White} D.~A.,    {Rangarajan} F.~V.~N.,  2006, \mnras, 368, 65

\bibitem[\protect\citeauthoryear{{Ellis} \& {Jones}}{{Ellis} \&
  {Jones}}{2004}]{ell04}
{Ellis} S.~C.,  {Jones} L.~R.,  2004, \mnras, 348, 165

\bibitem[\protect\citeauthoryear{{Ellis}, {Jones}, {Donovan}, {Ebeling} \&
  {Khosroshahi}}{{Ellis} et~al.}{2006}]{ell06}
{Ellis} S.~C.,  {Jones} L.~R.,  {Donovan} D.,  {Ebeling} H.,    {Khosroshahi}
  H.~G.,  2006, \mnras, 368, 769

\bibitem[\protect\citeauthoryear{{Ettori}}{{Ettori}}{2005}]{ett05}
{Ettori} S.,  2005, \mnras, 362, 110

\bibitem[\protect\citeauthoryear{{Ettori}, {Tozzi} \& {Rosati}}{{Ettori}
  et~al.}{2003}]{ett03}
{Ettori} S.,  {Tozzi} P.,    {Rosati} P.,  2003, \aap, 398, 879

\bibitem[\protect\citeauthoryear{{Fukazawa}, {Makishima}, {Tamura}, {Ezawa},
  {Xu}, {Ikebe}, {Kikuchi} \& {Ohashi}}{{Fukazawa} et~al.}{1998}]{fuk98a}
{Fukazawa} Y.,  {Makishima} K.,  {Tamura} T.,  {Ezawa} H.,  {Xu} H.,  {Ikebe}
  Y.,  {Kikuchi} K.,    {Ohashi} T.,  1998, \pasj, 50, 187

\bibitem[\protect\citeauthoryear{{Henry}}{{Henry}}{2004}]{hen04}
{Henry} J.~P.,  2004, \apj, 609, 603

\bibitem[\protect\citeauthoryear{{Jee}, {White}, {Ford}, {Blakeslee},
  {Illingworth}, {Coe} \& {Tran}}{{Jee} et~al.}{2005}]{jee05b}
{Jee} M.~J.,  {White} R.~L.,  {Ford} H.~C.,  {Blakeslee} J.~P.,  {Illingworth}
  G.~D.,  {Coe} D.~A.,    {Tran} K.-V.~H.,  2005, \apj, 634, 813

\bibitem[\protect\citeauthoryear{{Jeltema}, {Canizares}, {Bautz}, {Malm},
  {Donahue} \& {Garmire}}{{Jeltema} et~al.}{2001}]{jel01}
{Jeltema} T.~E.,  {Canizares} C.~R.,  {Bautz} M.~W.,  {Malm} M.~R.,  {Donahue}
  M.,    {Garmire} G.~P.,  2001, \apj, 562, 124

\bibitem[\protect\citeauthoryear{Joy, LaRoque, Grego, Carlstrom, Dawson,
  Ebeling, Holzapfel, Nagai \& Reese}{Joy et~al.}{2001}]{joy01}
Joy M.,  LaRoque S.,  Grego L.,  Carlstrom J.~E.,  Dawson K.,  Ebeling H.,
  Holzapfel W.~L.,  Nagai D.,    Reese E.~D.,  2001, \apj, 551, L1

\bibitem[\protect\citeauthoryear{Kaastra \& Mewe}{Kaastra \&
  Mewe}{1993}]{kaa93}
Kaastra J.~S.,  Mewe R.,  1993, {\aaps}, 97, 443

\bibitem[\protect\citeauthoryear{{Khosroshahi}, {Maughan}, {Ponman} \&
  {Jones}}{{Khosroshahi} et~al.}{2006}]{kho06}
{Khosroshahi} H.~G.,  {Maughan} B.~J.,  {Ponman} T.~J.,    {Jones} L.~R.,
  2006, \mnras, pp 514--+

\bibitem[\protect\citeauthoryear{{Kravtsov}, {Vikhlinin} \& {Nagai}}{{Kravtsov}
  et~al.}{2006}]{kra06a}
{Kravtsov} A.~V.,  {Vikhlinin} A.,    {Nagai} D.,  2006, ArXiv Astrophysics
  e-prints

\bibitem[\protect\citeauthoryear{{Majumdar} \& {Mohr}}{{Majumdar} \&
  {Mohr}}{2003}]{maj03}
{Majumdar} S.,  {Mohr} J.~J.,  2003, \apj, 585, 603

\bibitem[\protect\citeauthoryear{Markevitch}{Markevitch}{1998}]{mar98a}
Markevitch M.,  1998, \apj, 504, 27

\bibitem[\protect\citeauthoryear{Markevitch}{Markevitch}{2001}]{mar00b}
Markevitch M.,  2001, ACIS background.
http://hea-www.harvard.edu/~maxim/axaf/acisbg/

\bibitem[\protect\citeauthoryear{{Mathiesen} \& {Evrard}}{{Mathiesen} \&
  {Evrard}}{2001}]{mat01b}
{Mathiesen} B.~F.,  {Evrard} A.~E.,  2001, \apj, 546, 100

\bibitem[\protect\citeauthoryear{{Maughan}, {Ellis}, {Jones}, {Mason},
  {Cordova} \& {Priedhorsky}}{{Maughan} et~al.}{2006}]{mau06b}
{Maughan} B.~J.,  {Ellis} S.~C.,  {Jones} L.~R.,  {Mason} K.~O.,  {Cordova} F.,
     {Priedhorsky} W.~C.,  2006, \apj, in press

\bibitem[\protect\citeauthoryear{{Maughan}, {Jones}, {Ebeling} \&
  {Scharf}}{{Maughan} et~al.}{2004}]{mau04a}
{Maughan} B.~J.,  {Jones} L.~R.,  {Ebeling} H.,    {Scharf} C.,  2004, \mnras,
  351, 1193

\bibitem[\protect\citeauthoryear{{Maughan}, {Jones}, {Ebeling} \&
  {Scharf}}{{Maughan} et~al.}{2006}]{mau06a}
{Maughan} B.~J.,  {Jones} L.~R.,  {Ebeling} H.,    {Scharf} C.,  2006, \mnras,
  365, 509

\bibitem[\protect\citeauthoryear{{Maughan}, {Jones}, {Lumb}, {Ebeling} \&
  {Gondoin}}{{Maughan} et~al.}{2004}]{mau04b}
{Maughan} B.~J.,  {Jones} L.~R.,  {Lumb} D.,  {Ebeling} H.,    {Gondoin} P.,
  2004, \mnras, 354, 1

\bibitem[\protect\citeauthoryear{{Navarro}, {Frenk} \& {White}}{{Navarro}
  et~al.}{1996}]{nav96}
{Navarro} J.~F.,  {Frenk} C.~S.,    {White} S.~D.~M.,  1996, \apj, 462, 563

\bibitem[\protect\citeauthoryear{{Navarro}, {Frenk} \& {White}}{{Navarro}
  et~al.}{1997}]{nav97}
{Navarro} J.~F.,  {Frenk} C.~S.,    {White} S.~D.~M.,  1997, \apj, 490, 493

\bibitem[\protect\citeauthoryear{{O'Sullivan}, {Vrtilek}, {Kempner}, {David} \&
  {Houck}}{{O'Sullivan} et~al.}{2005}]{osu05}
{O'Sullivan} E.,  {Vrtilek} J.~M.,  {Kempner} J.~C.,  {David} L.~P.,    {Houck}
  J.~C.,  2005, \mnras, 357, 1134

\bibitem[\protect\citeauthoryear{{Ponman}, {Cannon} \& {Navarro}}{{Ponman}
  et~al.}{1999}]{pon99}
{Ponman} T.~J.,  {Cannon} D.~B.,    {Navarro} J.~F.,  1999, \nat, 397, 135

\bibitem[\protect\citeauthoryear{{Randall}, {Sarazin} \& {Ricker}}{{Randall}
  et~al.}{2002}]{ran02}
{Randall} S.~W.,  {Sarazin} C.~L.,    {Ricker} P.~M.,  2002, \apj, 577, 579

\bibitem[\protect\citeauthoryear{{Read} \& {Ponman}}{{Read} \&
  {Ponman}}{2003}]{rea03}
{Read} A.~M.,  {Ponman} T.~J.,  2003, \aap, 409, 395

\bibitem[\protect\citeauthoryear{{Rowley}, {Thomas} \& {Kay}}{{Rowley}
  et~al.}{2004}]{row04}
{Rowley} D.~R.,  {Thomas} P.~A.,    {Kay} S.~T.,  2004, \mnras, 352, 508

\bibitem[\protect\citeauthoryear{{Sarazin}}{{Sarazin}}{1986}]{sar86}
{Sarazin} C.~L.,  1986, Reviews of Modern Physics, 58, 1

\bibitem[\protect\citeauthoryear{Scharf, Jones, Ebeling, Perlman, Malkan \&
  Wegner}{Scharf et~al.}{1997}]{sch97}
Scharf C.,  Jones L.~R.,  Ebeling H.,  Perlman E.,  Malkan M.,    Wegner G.,
  1997, \apj, 477, 79

\bibitem[\protect\citeauthoryear{{Smith}, {Brickhouse}, {Liedahl} \&
  {Raymond}}{{Smith} et~al.}{2001}]{smi01}
{Smith} R.~K.,  {Brickhouse} N.~S.,  {Liedahl} D.~A.,    {Raymond} J.~C.,
  2001, \apjl, 556, L91

\bibitem[\protect\citeauthoryear{{Spergel}, {Verde}, {Peiris}, {Komatsu},
  {Nolta}, {Bennett}, {Halpern}, {Hinshaw}, {Jarosik}, {Kogut}, {Limon},
  {Meyer}, {Page}, {Tucker}, {Weiland}, {Wollack} \& {Wright}}{{Spergel}
  et~al.}{2003}]{spe03}
{Spergel} D.~N.,  {Verde} L.,  {Peiris} H.~V.,  {Komatsu} E.,  {Nolta} M.~R.,
  {Bennett} C.~L.,  {Halpern} M.,  {Hinshaw} G.,  {Jarosik} N.,  {Kogut} A.,
  {Limon} M.,  {Meyer} S.~S.,  {Page} L.,  {Tucker} G.~S.,  {Weiland} J.~L.,
  {Wollack} E.,    {Wright} E.~L.,  2003, \apjs, 148, 175

\bibitem[\protect\citeauthoryear{{Tonry}, {Schmidt} \& et. al.}{{Tonry}
  et~al.}{2003}]{ton03}
{Tonry} J.~L.,  {Schmidt} B.~P.,    et. al. 2003, \apj, 594, 1

\bibitem[\protect\citeauthoryear{{Vikhlinin}}{{Vikhlinin}}{2006}]{vik06b}
{Vikhlinin} A.,  2006, \apj, 640, 710

\bibitem[\protect\citeauthoryear{{Vikhlinin}, {Kravtsov}, {Forman}, {Jones},
  {Markevitch}, {Murray} \& {Van Speybroeck}}{{Vikhlinin}
  et~al.}{2006}]{vik06a}
{Vikhlinin} A.,  {Kravtsov} A.,  {Forman} W.,  {Jones} C.,  {Markevitch} M.,
  {Murray} S.~S.,    {Van Speybroeck} L.,  2006, \apj, 640, 691

\bibitem[\protect\citeauthoryear{{Vikhlinin}, {Markevitch}, {Murray}, {Jones},
  {Forman} \& {Van Speybroeck}}{{Vikhlinin} et~al.}{2005}]{vik05a}
{Vikhlinin} A.,  {Markevitch} M.,  {Murray} S.~S.,  {Jones} C.,  {Forman} W.,
   {Van Speybroeck} L.,  2005, \apj, 628, 655

\bibitem[\protect\citeauthoryear{{Vikhlinin}, {McNamara}, {Forman}, {Jones},
  {Quintana} \& {Hornstrup}}{{Vikhlinin} et~al.}{1998}]{vik98b}
{Vikhlinin} A.,  {McNamara} B.~R.,  {Forman} W.,  {Jones} C.,  {Quintana} H.,
   {Hornstrup} A.,  1998, \apj, 502, 558

\bibitem[\protect\citeauthoryear{{Vikhlinin}, {Voevodkin}, {Mullis},
  {VanSpeybroeck}, {Quintana}, {McNamara}, {Gioia}, {Hornstrup}, {Henry},
  {Forman} \& {Jones}}{{Vikhlinin} et~al.}{2003}]{vik03}
{Vikhlinin} A.,  {Voevodkin} A.,  {Mullis} C.~R.,  {VanSpeybroeck} L.,
  {Quintana} H.,  {McNamara} B.~R.,  {Gioia} I.,  {Hornstrup} A.,  {Henry}
  J.~P.,  {Forman} W.~R.,    {Jones} C.,  2003, \apj, 590, 15

\bibitem[\protect\citeauthoryear{{Voit} \& {Donahue}}{{Voit} \&
  {Donahue}}{2005}]{voi05b}
{Voit} G.~M.,  {Donahue} M.,  2005, \apj, 634, 955

\bibitem[\protect\citeauthoryear{{Voit}, {Kay} \& {Bryan}}{{Voit}
  et~al.}{2005}]{voi05c}
{Voit} G.~M.,  {Kay} S.~T.,    {Bryan} G.~L.,  2005, \mnras, 364, 909

\bibitem[\protect\citeauthoryear{{Wechsler}, {Bullock}, {Primack}, {Kravtsov}
  \& {Dekel}}{{Wechsler} et~al.}{2002}]{wec02}
{Wechsler} R.~H.,  {Bullock} J.~S.,  {Primack} J.~R.,  {Kravtsov} A.~V.,
  {Dekel} A.,  2002, \apj, 568, 52

\end{thebibliography}

\end{document}